\documentclass[aoas,preprint]{imsart}

\RequirePackage[OT1]{fontenc}
\RequirePackage{amsthm,amsmath,amsfonts,natbib,graphicx}
\RequirePackage[colorlinks]{hyperref}
\RequirePackage{hypernat}

\startlocaldefs
\numberwithin{equation}{section}
\theoremstyle{plain}
\newtheorem{theorem}{Theorem}[section]
\newtheorem{proposition}{Proposition}[section] 
\newtheorem{lemma}{Lemma}[section] 
\newtheorem{assumption}{Assumption} 
\newtheorem{remark}{Remark}[section]

\endlocaldefs
\begin{document}

\begin{frontmatter}
\title{Inference for partially observed multitype branching processes
  and  ecological applications}
\runtitle{Partially observed multitype processes}
\begin{aug}
\author{\snm{Catherine Lar\'edo }\thanksref{t1}
\ead[label=e1]{catherine.laredo@jouy.inra.fr}},
\author{
\snm{Olivier David }
\ead[label=e2]
{olivier.david@jouy.inra.fr}}, 

\and
\author{
\snm{Aur\'elie Garnier}`
\ead[label=e3]
{au.garnier@gmail.com}}

\thankstext{t1}{First supporter of the project}

\runauthor{C. Lar\'edo  et al.}

\affiliation{INRA, Jouy-en-Josas and PMA Universit\'es Paris 6 \& 7}

\address{Catherine Lar\'edo,\\
UR341, Math\'ematiques et 
Informatique appliqu\'ees,\\
INRA, F-78350 Jouy-en-Josas, France \\
\& CNRS, UMR7599, Probabilit\'es et Mod\`eles al\'eatoires,\\
Universit\'es Paris 6 \& 7,  75005 Paris France.\\
\printead{e1}}


\address{Olivier David,\\ 
UR341, Math\'ematiques et\\ 
Informatique appliqu\'ees,\\
INRA, F-78350 Jouy-en-Josas,France.\\
\printead{e2}}\

\address{Aur\'elie Garnier \\
Universit\'e Paris Sud, \\
Laboratoire Ecologie, Syt\'ematique \& Evolution,\\ 
UMR8079, F-91405 Orsay, France;CNRS, 91405 0rsay, France; AgroParisTech, 75231 Paris, France.\\
\printead{e3}} \

\end{aug}

\begin{abstract}
	Multitype branching processes with immigration in one
 type are used to model the dynamics of stage-structured plant 
populations. Parametric inference is first carried out 
when count data of all types are observed. Statistical 
identifiability is proved together with derivation of consistent 
and asymptotically Gaussian estimators for all the parameters ruling 
the population dynamics model. However, for many ecological data, 
some stages (i.e. types) cannot be observed in practice. 
We study which mechanisms can still be estimated 
given the model and the data available in this context. 
Parametric inference is investigated in the case of 
Poisson distributions. We prove that identifiability holds 
for only a subset of the parameter set depending on  the number of generations
observed, together with consistent 
and asymptotic properties of estimators. 
Finally, simulations 
are performed to study the behaviour of the estimators 
when the model is no longer Poisson. Quite good results are obtained 
for a large class of models with distributions having mean 
and variance within the same order of magnitude, leading to 
some stability results with respect to the Poisson assumption.
\end{abstract}

\begin{keyword}[class=AMS]
\kwd[Primary ]{62M05, 62M09, 62P12}
\kwd[; secondary ]{92D40,60J85}
\end{keyword}

\begin{keyword}
\kwd{Parametric inference, Incomplete data, Multitype branching processes,
ecology.}
\end{keyword}

\end{frontmatter}

\section{Introduction} \label{sect:intro}

Understanding population dynamics requires models that admit the
complexity of natural populations and the data ecologists 
can get from them. Thus analyzing ecological data  raises 
questions ranging from modeling purposes to statistical inference.
Among various methods, Leslie matrices or demographic
matrix models  are widely 
used for studying the dynamics of 
age or stage-structured populations (e.g. \citealp{Caswell:2001}). 
These models are deterministic with noise added to introduce some variability. 
In many cases however and  especially for small populations, 
the demographic stochasticity has to be taken into account; 
these models are too simple 
\citep{Melbourne:2008} and can no longer be used, even
adding stochasticity 
into the dynamics with random effects and  
covariates (\citealp{Royle:2004, Barry:2003}) or 
Bayesian approaches (\citealp{Raftery:1995, Gross:2002, Clark:2004}).
For these reasons, we use here
stochastic models to study small populations dynamics. 

The starting point of this work is a three-year field survey of feral
populations (i.e. populations escaped from crops) of an annual crop species
(oilseed rape) that was carrried out in the center of France 
(Selommes, Loir-et-Cher; \citealp{Garnier:2006c}).
Unlike cultivated oilseed rape, very few facts are known 
about the dynamics of feral oilseed rape populations.
In this study, the  dynamics is modeled by a multitype
 branching process (five types including vegetative and reproductive 
plant stages along with seeds in the soil seedbank) with immigration in 
one type (seeds).
Data consisted  in populations counts in each type, except the seeds
that could not be observed.
Three main  difficulties occur when studying 
this demographic dataset. (1) A large number  
of populations ($K=300$)
have been observed over a short period of time ($n=2,3$);
(2) only count data have been collected;
(3) some types  could not be observed by ecologists (here seeds).     
These characteristics are clearly not specific 
to this survey  and are frequently met
in data coming from Population Genetics
and Ecology (see e.g. \citealp{deValpine:2004} and the references therein). 
These data could be studied as longitudinal data, but for concerns about 
the dynamics, better insights can be obtained by means of mechanistic
models describing it. 

Branching processes have largely been studied (see 
\citealp{Athreya:1972} for a general presentation;
\citealp{Mode:1971} for Multitype Branching Processes and 
\citealp{Haccou:2005, Kimmel:2002, Mode:2000} 
for applications in biology). Statistical
inference has also been largely investigated (\citealp{Hall:1980, Guttorp:1991} 
for general branching processes; 
\citealp{Wei:1990, Winnicki:1991} for branching processes
 with immigration;  
\citealp{Bhat:1981, Maaouia:2005, Gonzalez:2008} for multitype branching 
processes).
However, the precise multitype branching process 
with immigration used here is a combination of 
the previous ones and moreover statistical inference
for multitype branching 
processes is usually based 
on the following different observations (\citealp{Maaouia:2005,Gonzalez:2008}):
the number of descendants of type $j$ 
coming from all the type  $i$ individuals. 
We just observed the successive counts of "individuals" of each type. This is more realistic assumption 
since this situation frequently  occurs with datasets 
from field studies, and inference is studied here in this framework.

We are interested in the estimation of the parameters involved in  
the population dynamics from the incomplete observations of
count data collected simultaneously
in several populations. 
 This is an ``Incomplete Data model'',
or  ``State Space model'' 
(as defined for instance in \citep{Cappe:2005}.
It is also an inverse problem and   
a central theme in Ecology arising from its study is that parameters 
might not be identifiable 
knowing only the population dynamics (\citealp{Wood:1997}). 
In practice, the inference based on such data is performed using various
E.M. algorithms eventually coupled with 
Monte Carlo methods (\citealp{Dempster:1977, Kuhn:2004,
McLachlan:2007, Sung:2007, Olsson:2008}), and Bayesian or 
Hierarchical Bayesian methods   
(\citealp{Clark:2004,
Buckland:2004, Thomas:2005}). All these methods circumvent   
but cannot address the identifiability problem. 
However, identifiability is a prerequisite of 
statistical inference, and understanding the dynamics mechanisms 
strongly relies on how parameters are linked in the identifiability 
question.
We propose here an integrated framework in order 
to analyze as accurately as possible the whole data set of 
the field survey. Introduction 
of covariates and a priori knowledge, errors coming from 
non exhaustive population samplings 
within some populations, use of various algorithms rely on this work 
and are studied in two companion papers (\citealp{David:2008,
Garnier:2008}).

  The paper is organized as follows. Section \ref{sect:mod} contains 
the description of the population dynamics and preliminary results
(Proposition \ref{prop:Markov}).
The statistical inference for complete observations 
is  studied in Section \ref{sect:licomp}. We first prove
identifiability for all the parameters  and 
derive consistent and asymptotically Gaussian estimators 
(Proposition \ref{prop:mle}). 
Since seeds are not observed in practice,
the problem of unobserved types is addressed in Section 
\ref{sect:incomp}. This is a non linear non Gaussian state space model.
The associated three-dimensional stochastic process 
is no longer Markovian. 
The model with Poisson distributions provides 
a useful example with explicit computations. We obtain a closed form of the 
dependence of present observations 
on the whole past for the three-dimensional process 
(Theorem \ref{theo:cond}).
A question concerns the statistical model identifiability : 
it is  studied 
according to the number of observed generations.
We characterize the parameter subset where identifiability holds 
(Theorem \ref{theo:identifiability}) 
and study the parametric inference (Proposition  \ref{prop:statphi}). 
Section \ref{sect:simul}
presents simulation results to study how the estimation performs 
with respect to deviations to the Poisson model. Detailed proofs are
given in the Appendix.

\section{ Model and preliminary results}\label{sect:mod}
\subsection{Dynamics of annual plants} \label{sub:dyn}
We consider annual plants with the following life cycle.
Seeds are released at the end of summer; they can either 
enter in a seed bank if buried
or germinate in autumn.The emerged rosettes  vernalize during winter, 
then bolt in spring and finally produce mature plants 
that shed seeds in summer and then die. Five developmental 
stages are considered: rosettes before winter $R$, rosettes after
winter (vernalized rosettes)
 $V$, mature plants $F$, seeds located in  
the soil seed bank (``old
seeds'') $S$, and seeds located on the soil surface (``new
seeds'') $T$. New seeds and old seeds are separated 
because they have different 
demographic parameters. Within each cycle, new seeds can enter 
these populations at the end of summer
(immigration). There exist two sources of seed immigration, seeds from
adjacent mature crops and seeds from spillage  during seed transport
(\citealp{Crawley:1995, Claessen:2005a, Garnier:2006a, Pivard:2008}).

\bigskip 

\begin{figure}[htp]
\centering
\includegraphics[height=10cm]{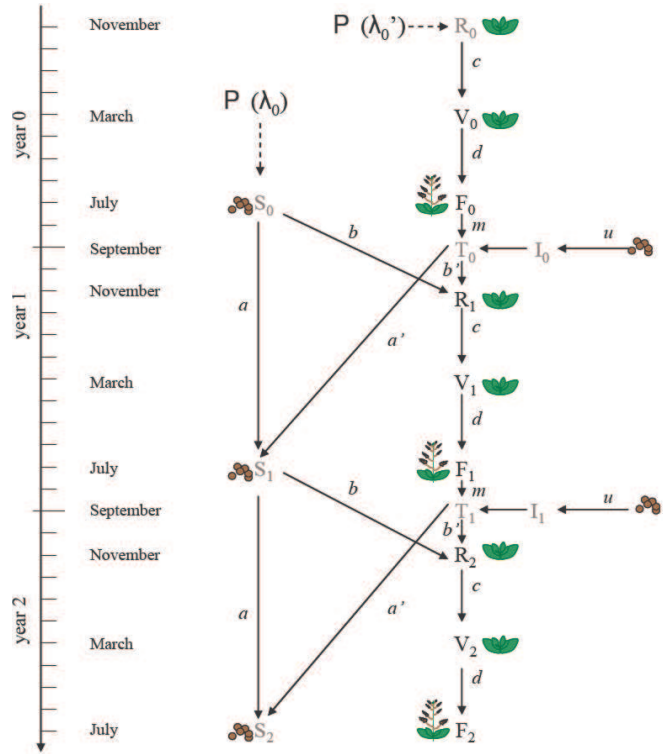}
\caption{Schematic Dynamics of feral oilseed rape populations.}
\end{figure}

\bigskip

This  model is quite general and for dynamical purposes only, it could  
be simplified
considering just seeds and mature plants. However, our concern is
different since we aim at estimating as many parameters as
possible given both the model and the data available.
Keeping a five-type model allows using  all the 
data collected in the field survey and thus leads to the best description 
of the plant population dynamics for  inference.

\subsection{Notations and assumptions} \label{sub:not}
Consider first one population.
From now on, the term year corresponds to one life cycle. It  
starts with the birth of
the new seeds and ends just before the birth of the new seeds of 
the next generation.
All the variables are integer random variables 
indexed by $i \in \mathbb{N}$.
For  year $i$, denote by $S_i$  the number of "old seeds", 
 $T_i$ the
number of "new seeds", $R_i$ the number of 
rosettes before winter, $V_i$ the number of rosettes after 
winter and $F_i$ the
number of mature plants.
Six parameters describe these transitions: 
\begin{equation}\label{def:cda}
(c,d,a,b,a',b') \in (0,1)^6 \mbox { with }
\;0< a+b \leq 1,\; 0<a'+b'\leq 1.
\end{equation}
\begin{equation}\label{def:ab}
P( \mbox{seed in } S_i \rightarrow 
S_{i+1})= a \;;\;
P( \mbox{seed in } S_i \rightarrow
R_{i}) = b             
\end{equation}
\begin{equation}\label{def:a'b'}
P( \mbox{seed in }T_i \rightarrow 
S_{i+1}) = \;a'\; ;\;
P( \mbox{seed in } T_i \rightarrow 
R_{i})= \;b'
\end{equation}
\begin{equation}\label{def:cd} 
 P( \mbox{rosette in } R_i \rightarrow 
V_i ) =  \;c\;;\;
 P( \mbox{rosette in } V_i \rightarrow 
F_i)= \;d.
\end{equation} 
Mature plants in $F_{i}$ produce ``new seeds'' $T'_{i+1}\;$  
according to the offspring distribution $G(.)$.

A number $I_i$ of "new seeds" immigrate into the population
at the beginning of year $i$; it is assumed to follow the distribution $\mu(.)$.
Seeds in $S_i$ come from two sources, the "old seeds"  $S_{i-1}$
 and the "new seeds"  $T_{i-1}$. Denote by  $S'_i$ and  $S''_i$
 these two quantities.
Similarly, rosettes before winter $R_i$ come from ``old seeds'' in
$S_i$  and ``new seeds'' in $T_i$. Denote by  
$R'_{i}$ and $R''_{i}$ the rosettes coming from $S_{i}$ and
$T_{i}$. These variables satisfy :
\begin{equation}\label{def:S'R'}
S_i=S'_i+S''_i\; \; , \;\; R_i=R'_i+R''_i \;,
\end{equation} 
\begin{equation}\label{def:T}
T_i=T'_i+I_i\;.
\end{equation} 
Note that the  probabilities of dying for stages $S, T, R, V$ are 
respectively $(1-a -b)$,
$(1-a'-b')$, $(1-c)$, $(1-d)$ and that the probability of 
no offspring for a mature plant is $G(0)$.

Let us now detail the framework and assumptions used in the sequel.
The field survey consisted of a large number of feral oilseed rape 
populations (around 300 in \citealp{Garnier:2006c}) 
observed over a short period of time ($n$ = 2 or 3).
 These populations were isolated, so  we assume here independence for 
these populations and the plant density was low in the surveyed
populations so that density-dependence in plant 
survival and reproduction could be neglected
(\citealp{Pivard:2008, Garnier:2006c}). Moreover, we assume
\begin{assumption} \label{G:iid}
Offspring distribution $G(.)$\\ 
All mature plants reproduce independently according to the same 
offspring distribution $G(.)$ with expectation $m$ and variance $\delta^2$.
\end{assumption}
\begin{assumption}\label{SiTi} 
There is no competition in survival and germination 
 between seeds in the seed bank or 
``old seeds'' $S_i$ and seeds on the soil or ``new seeds'' $T_i$.
\end{assumption}
\begin{assumption}\label{a:tind}
Immigration distribution $\mu(.)$\\
(i) Immigration $I_i$ is independent of seed bank seeds $S_i$,
offspring seeds $T'_i$  and of previous years.\\
(ii) The random variables $(I_i, i \in N)$ are independent and identically
distributed according to  $\mu(.) $ with
expectation $u$ and variance $\rho^2$.
\end{assumption}

The above assumptions are twofold: ``independence'' and
``identically distributed''. They do not have the same status :
while releasing the ``independence '' assumption is quite difficult, 
the ``identically distributed '' 
assumption is done here for sake of clarity.   
The independence assumption in (A\ref{G:iid}) is justified because of the low 
 plant density.
There is no biological background for considering
competition between the evolution of "old seeds"  $S_i$ and "new seeds" $T'_i$
leading to (A\ref{SiTi}). In the field survey of feral 
 oilseed rape populations, seed immigration mainly occured from spillage during seed transport and from adjacent mature crops, leading to the independence
assumptions in (A\ref{a:tind})
(\citealp{Pivard:2008, Garnier:2006c}).
Adding covariates 
or a priori information  can easily be introduced within this 
framework, which 
amounts to remove the ``identically distributed'' assumption.
This indeed has been done  for the statistical
 analysis on the whole data set: using that many populations had been
 observed, covariates, a priori knowledge, and  
a dependence with respect to  $i,k$  of the parameters
defined in (\ref{def:ab})-(\ref{def:cd})
have been added 
(\citealp{Garnier:2008}). 
The framework detailed here is therefore quite general.

\subsection{Preliminary results}\label{sub:prelim}
From now on, time $i$ is 
associated with a complete cycle of a plant. 
Let us still consider one population.  
The model is a discrete time stochastic 
process $(X_{i})$ with state space 
${\mathbb{N}}^5$. Set 
\begin{equation}\label{def:X} 
X_{i}=(S_{i},T_i, R_{i},V_{i,}F_{i}) \;\mbox{and }  
{\cal F}_i= \sigma (X_0,X_1,\dots,,X_{i}). 
\end{equation}
For $x=(s,t,r,v,f)$, $x'=(s',t', r',v',f')
\in \mathbb{N}^5$, 
denote $\pi _{0} (x)$ the initial distribution 
and $p(x,x')$ the conditional distribution of $X_1$ given $X_0$ : 
\begin{equation}\label{def:trans} 
	\pi_{0} (x) =
\mathbb{P}( X_{0}= x) \;\;;\;\; p(x,x') = 
	\mathbb{P}(X_{1}=x'/ X_{0}= x). 
\end{equation}
For the Binomial distribution ${\cal B} (N;p)$,
 we write $\mathbb{P}(Y=k)={\cal B} (N;p)(k)$;
Multinomial distributions on $\mathbb{ N}^l $,
 ${\cal M} (N;p_1,\ldots,p_l)$ 
are simplified omitting the last component, which leads for $l=3$ and
$0<p_1+p_2<1$, 
\begin{equation}\label{def:multi} 
	{\cal M} (N;p_1,p_2)(n_1,n_2)=
{\cal M} (N;p_1,p_2,1-p_1 -p_2)(n_1,n_2,N-n_1-n_2)
\end{equation}
Let $\star$ denote the convolution product of two distributions. 
Using these notations 
and (\ref{def:cda}), 
define the distributions 
\begin{equation}\label{def:probinit} 
	\nu (s,t) = \mathbb{P}(S_{0} =s, T_{0}=t) , 
\end{equation}
\begin{equation}\label{def:p1}
	p_1(s'/s,t,r)= \frac{({\cal M}(s; a,b) \star 
	{\cal M}(t;a',b'))(s',r)} {({\cal B}(s;b)\star {\cal B}(t;b'))(r) }, 
\end{equation}
\begin{equation}\label{def:p2p3} 
	p_2(t'/f)=(G^{\star f} \star \mu )(t');\;\; 
	p_3(r/s,t)= ({\cal B}(s;b)\star {\cal B}(t;b'))(r), 
\end{equation}
\begin{equation}\label{def:p4p5} 
	p_4(v/r)= {\cal B}(r;c)(v) \;, \; p_5(f/v)={\cal B}(v;d)(f). 
\end{equation}
Using now definitions (\ref{def:X}), (\ref{def:trans})  
and notations (\ref{def:multi})- 
(\ref{def:p4p5}), the following holds.
\begin{proposition}\label{prop:Markov} 
Under (A\ref{G:iid}), (A\ref{SiTi}), (A\ref{a:tind}), 	
$(X_i)_{i\geq 0}$ is a time homogeneous Markov chain on 
$\mathbb{N}^5$ 
with initial distribution $\pi _{0} (x)$ and 
transition probabilities $p(x,x')$ satisfying 
\begin{align}
\label{eq:pitrans} \pi _{0} (x) &= \nu (s,t) 
\;p_3(r/s,t) \; p_4(v/r)\; p_5(f/v)\\
p(x,x') &= p_1(s'/s,t,r)\; p_2(t'/f) \; p_3(r'/s',t') 
		\; p_4(v'/r') \; p_5(f'/v'). 
\end{align}
The process $(X_i)_{i\geq 0}$ is also a multitype branching 
process with immigration in one type.
\end{proposition}
 The last statement of Proposition \ref{prop:Markov}
is immediate since, considering each stage of the plant as a type, 
each plant reproduces independently from the others according to the same 
offspring distribution with values in $\mathbb{N}^5$. 
However, the two types $R_i$ and $V_i$ have no offspring 
in the next generation, leading to
a non positively regular multitype process as defined in \cite{Mode:1971} or 
\cite{Athreya:1972}. The process $(S_i,T_i,F_i)$ is positively regular 
process but, for the reasons stated in \ref{sub:dyn}, we prefer keeping  here 
the five-dimensional process $(X_i)$.

The proof of Proposition \ref{prop:Markov} is given in Appendix 
\ref{appen:Markov}.

\section{Likelihood and inference for complete observations}
\label{sect:licomp}
We first study the case when all types are observed.
\subsection{Notations and statistical framework}\label{sub:linot}
We assume in the sequel that
the initial distribution of $(S_0,R_0)$, 
the offspring distribution $G(.)$ and the immigration 
$\mu(.)$ belong to the parametric families :\\
- distribution of $(S_0,T_0)$ : 
$\big ( \nu(\theta ^1;s,t),\;\theta ^1 \in \Theta ^1 \big);$\\
- offspring $G(.)$: $ \big( G(\theta^{2};.)\;, 
\;\theta ^2 \in \Theta ^2 \big)$ with mean $m$ and variance $\delta^2$;\\   
- immigration $\mu(.) $ : 
$\big( \mu (\theta^{3};.), \;\theta ^3 \in \Theta ^3)$ with mean $u$ 
and variance $\rho^2$.\\

Let us denote by $
\theta =(c,d,a,b,a',b',\theta^{1},
\theta^{2}, \theta^{3})$ (resp. $\theta_0$)
 an arbitrary value (resp. the true value) of the parameter and by $\Theta$
 the parameter set. We assume :
 
\begin{assumption}\label{ass:param}
$\Theta$ compact set of $\mathbb{R}^l$ and $\theta_0 \in 
\overset{\circ}{\Theta}$.
\end{assumption}

For the $k$th population, $X^{k}=(X_i^k, i\in 
\mathbb{N})$ is the Markov chain describing 
the population dynamics and $x_i^k = (s_i^k, t_i^k,r_i^k,v_i^k,f_i^k)$ are
the  observations at time $i$.
 In order to simplify the expressions for the likelihood, 
we consider here that $ S_i$, $T_i$ are observed 
up to time $n+1$. This has no consequence since seeds are not 
observed in practice.  
Hence we denote,
\begin{align}\label{def:obsk}
X^{k}_{0:n} &=\big ( X_0^k,\dots, X_n^k,  
S^k_{n+1},T^k_{n+1}\big ),\\
{\cal O}_{0:n}^k & = (x_0^k, x_1^k,\dots,x_n^k,s_{n+1}^k, t_{n+1}^k).
\end{align}
The processes $X^{k}_{0:n}$ are repetitions of 
$X_{0:n}=(X_0,\dots, X_n, S_{n+1},T_{n+1})$. 
Joining all 
the populations, define : 
\begin{equation}
	\label{def:Obser} 
X_{0:n}(K)= (X_{0:n}^{1},\dots,X_{0:n}^{K}),\;\;\mbox{ and  }\;\;  
{\cal O}_{0:n}(K)=({\cal O}_{0:n}^1,\dots,{\cal O}_{0:n}^ K). 
\end{equation}
Let $\pi_0(\theta;x)$ (resp. $p(\theta ; x,x')$) be the initial distribution
(resp. the transition probabilities) associated with parameter
$\theta$ and $\mathbb{P}_{\theta}$ the probability distribution of
$(X_i)$ on the canonical space and $\mathbb{E}_{\theta}$ the expectation w.r.t.
$\mathbb{P}_{\theta}$.

\subsection{Likelihood}\label{sub:licomp}
Computing the likelihood of a Markov process with transition probabilities
 $p(\theta;x,x')$ is classical : 
for population $k$, 
it has for expression, using (\ref{eq:pitrans}), 
\begin{equation}
	\label{eq:LOnk} 
L(\theta; {\cal O}_{0:n}^k) = \pi_0(\theta;x_{0}^{k}) 
 	\bigl(\prod _{i=1}^{n} p(\theta; x_{i-1}^k,x_{i}^k)\bigr) 
p_1(s_{n+1}^k/x_{n}^k) p_2(t_{n+1}^k/x_{n}^k). 
\end{equation}
Joining the observations from the $K$ independent populations, we
obtain using Proposition \ref{prop:Markov}, (\ref{def:probinit})  -
(\ref{def:p4p5}) 
\begin{equation}\label{eq:lOn} 
l(\theta; {\cal O}_{0:n}(K)) = l_K(\theta)= 
\sum _{k=1}^{K} \log L(\theta; {\cal O}_{0:n}^k). 
\end{equation}
Reordering the terms of (\ref{eq:LOnk}) and (\ref{eq:lOn})
 according to the parameters yields,
\begin{equation}
	l(\theta;{\cal O}_{0:n}(K)) 
= \sum_{j=0}^{5} l_K^j (\theta).
\end{equation}
The first term deals with the initial distribution of $(S_0, T_0)$ 
\begin{equation}\label{eq:l0} 
l_K^0(\theta)
=\sum _{k=1}^{K} \log \nu (\theta^1; s_{0}^{k},t_{0}^{k}) 
=l_K^0(\theta ^1).
\end{equation}
Gather in the second term the transition from seeds $S,T$ to rosettes $R$. 
\begin{equation}\label{eq:l1} 
l_K^1(\theta)
	=\sum _{k=1}^{K} \sum _{i=0}^{n}\log ( {\cal B}(s_i^k;b) 
	\star {\cal B}(t_i^k;b'))(r_i^k)
	=l_K^1(b,b').
\end{equation}
The next two terms contain the transitions from $R$ to 
$V$ and $V$ to $F$,  they write
\begin{equation}\label{eq:l2} 
l_K^2(\theta)=
\sum _{k=1}^{K} \sum _{i=0}^{n}
 v_{i}^k \;\log c \; 
+ (r_{i}^k-v_{i}^k)\;\log (1-c) + C_2({\cal O}_{0:n}(K)) = l_{K}^2(c),
\end{equation}
\begin{equation}\label{eq:l3} 
l_K^3(\theta)=
\;\sum _{k=1}^{K} \sum _{i=0}^{n} 
f_{i}^k \log d +(v_{i}^k-f_{i}^k)\log (1-d)+C_3({\cal O}_{0:n}(K))
 = l_{K}^3(d),
\end{equation}
where the two terms $C_2({\cal O}_{0:n}(K))$ and $C_3({\cal O}_{0:n}(K))$ only depend on 
the observations. Set in the next term the transition from $F$ to $T$, 
\begin{equation}\label{eq:l4} 
l_K^4(\theta)=
\;\sum _{k=1}^{K} \sum _{i=0}^{n} \log \big(G(\theta^2;.)^{\star f_{i}^k} 
\star \mu (\theta^3;.)
	\big)(t_{i+1}^k) = l_{K}^4(\theta ^{2}, \theta ^{3})
\end{equation}
The last term concerns the seeds $S$ :  
\begin{equation}\label{eq:l5} 
l_K^5(\theta)=
\sum _{k=1}^{K}\sum _{i=0}^{n} \log(\frac{({\cal M}(s_{i}^k; a,b) 
\star {\cal M}(t_{i}^{k}; a',b'))(s_{i+1}^k,r_{i}^k)}
	{({\cal B}(s_i^k;b) \star {\cal B}(t_i^k;b')\big)(r_i^k)}) 
\end{equation}
Joining the two terms containing $a,a',b,b'$ yields
\begin{equation}\label{eq:l6}
l_K^6(\theta)
=  l_K^1(\theta) + l_K^5(\theta)
= l_K^6 (a,b,a',b').
\end{equation}

The terms $l_k^0(\theta),l_K^2(\theta),l_K^3(\theta),l_K^4(\theta)$ 
and $l_K^6(\theta)$ depend on disjoint sets 
for the parameters. Hence, maximizing the loglikelihood 
can be performed 
maximizing separately these five terms.
\begin{remark}\label{rem:largeK} 
Usually, statistical inference for Stochastic Processes is
investigated in the asymptotic framework small $K$ (mostly $K=1$) 
and large $n$ (leading to asymptotics results $n \rightarrow
+\infty$). Here, we have that $n$ is small 
 and $K$ large (e.g. magnitude 300).
This situation often occurs in Ecology (\citealp{deValpine:2004}). 
\end{remark}
\subsection{Study of maximum likelihood and other estimators}
This is a $K$ sample of i.i.d. random variables, each variable 
being a part of a branching process path. We have to use 
simultaneously the repetitions 
and the Markov structure to estimate the parameters. 
So, deriving the properties of the statistical model is 
not standard. 
The various terms $l_{K}^i(\theta)$ 
of the loglikelihood are associated with different parametric 
inference problems. 

The first term $l_{K}^0(\theta )$ deals with the estimation 
of $\theta ^1$ based on a sample of $K$ i.i.d. random variables 
with distribution $\nu (\theta ^1;.)$ on $\mathbb{N}^2$,
which is standard. The terms $l_{K}^2(\theta) = l_K^2( c)$
 and $l_{K}^3(\theta)= l_K^3(d) $ 
are related to the estimation of parameters $c, \;d$.
Denote by 
$\hat{c}_{K},\hat{d}_{K}$ the maximum likelihood  estimators (MLE)
obtained maximizing $l_K^2(c)$ and $l_K^3(d)$. They 
depend on the successive observations $(r_i^k,v_i^k)$
(resp. $(v_i^k,f_i^k)$) for $\{i=0,\ldots, n$; $ k=1,\ldots, K\}$
 and are explicit (see Appendix \ref{appen:licomp}).

Parameters $(a,b,a',b')$ are only present in 
$l_K^6(\theta)$ defined in (\ref{eq:l6}).
Maximum likelihood estimators for  $(a,b,a',b')$ can be defined maximizing 
$l_K^6(\theta)$. 
To prove identifiability and consistency, we have rather consider here 
conditional least squares (CLS)  estimators. 
Conditionally on $(S_i,T_i)$, 
the marginal distribution of $S_{i+1} $ (resp. $R_i$) is the sum of 
the two independent distributions, 
${\cal B}(S_i,a)$ and ${\cal B}(T_i,a')$ (resp. ${\cal B}(S_i,b)$ 
and $ {\cal B}(T_i,b')$). Therefore,
 we can define the CLS estimators $(\hat{a}_K,\hat{a}'_K)$ and 
$(\hat{b}_K,\hat{b}'_K)$  minimizing the Conditional Least Squares:
\begin{align}\label{def:J1J2}  
J_K^1(a,a') &= 
\sum _{k=1}^{K}\sum _{i=0}^{n}(s_{i+1}^k- a s_{i}^k -a't_{i}^k)^2\;;\; \\
J_K^2(b,b') &= 
\sum _{k=1}^{K}\sum _{i=0}^{n}(r_{i}^k- bs_{i}^k -b't_{i}^k)^2. 
\end{align}

The remaining term is $l_K^4(\theta^{2},\theta^{3}) = l_K^4(m,u)$
since we are concerned
by the estimation of $m$ and $u$.
This is the only part of the likelihood associated with 
the branching mechanism. 
This likelihood containing the convolution product $G^{\star{f}}\star
\mu$ is untractable,
and methods based on conditional least squares or weighted conditional 
least squares are used 
for branching processes (\citealp{Hall:1980,Guttorp:1991, Wei:1990})
leading to  moment estimations of $G$ and $\mu$. Noting 
that $E(T_{i+1}/F_i)= mF_i +u $, we can just consider for the
estimation of $m$ and $u$,
the CLS process,
\begin{equation}
	\label{def:J4} J_K^4 (m,u) =
\sum _{k=1}^{K}\sum _{i=0}^{n}(t_{i+1}^k-m f_i^k-u)^2. 
\end{equation}
Let  $(\hat{m}_K,\hat{u}_K)$ be the CLS estimators 
minimizing (\ref{def:J4}).
All the above estimators are explicitly defined in Appendix \ref{appen:licomp} 
and we can state :
\begin{proposition}\label{prop:mle} 
Assume (A\ref{G:iid}), (A\ref{SiTi}), (A\ref{a:tind}) and (A\ref{ass:param}). 
Then, under $\mathbb{P}_{\theta_0}$, 
all the parameters $c,d,a,b,a',b',m,u$ are identifiable and,
as $K\rightarrow \infty$,\\
{\bf(i)} $\;(\hat{c}_K,\hat{d}_K,\hat{a}_K,\hat{b}_K,\hat{a'}_K,\hat{b'}_K,
\hat{m}_K, \hat{u}_K)$ are consistent
and asymptotically Gaussian 
at rate $\sqrt{ K}$; \\
{\bf(ii)} $\; \hat{c}_K$, $\hat{d}_K$, $(\hat{a}_K,\hat{b}_K,
\hat{a'}_K,\hat{b'}_K)$,
$(\hat{m}_K, \hat{u}_K)$ 
are asymptotically independent, with explicit  covariance matrix 
 given in Appendix \ref{appen:licomp}.
\end{proposition}

Let us stress that, before studying in detail this inference problem,
it was difficult to assert the classical
properties stated in  Proposition \ref{prop:mle}. Adding immigration 
could lead to non identifiability or estimating problems for 
$m$ and $u$. Moreover, maximum likelihood estimators,
for multitype branching processes, 
are based on the observations of $G_{i,j}(k)$, i.e. offspring of type
$j$ from type $i$ parents (see \citealp{Guttorp:1991, Gonzalez:2008,
Maaouia:2005}). We did not require this information for the 
inference and just used the counts of individuals 
of each type in successive generations. Hence, getting identifiability and 
consistency for the parameters is the only difficulty here.
Other properties 
are classical but requires using the exact structure of the data
provided the regularity of the statistical model.

\begin{remark}\label{rem:estvar} Estimating additional moments of
  $G(.)$ and $\mu(.)$ can be performed similarly using other functionals
than Conditional Least Squares  (see \citealp{Winnicki:1991} for the
variance estimation of $G(.)$ and $\mu(.)$ ).
\end{remark}
The proof is given in Appendix \ref{appen:licomp}.

\section{Incomplete model study in the Poisson case} 
\label{sect:incomp}
The set-up is now different: only $R_i,V_i,F_i$ are observed 
while $S_i$ and $T_i$ are unobserved.
Clearly, algorithms simulating the missing data given the model 
and the parameters at each step can be used to get estimation. 
This approach is complementary to our concern that aims at understanding 
which mechanisms can be estimated.
For this, we have to study the process $(R_{i},V_{i},F_{i})$ 
for $ i=0,\ldots,n$. 
It is a discrete time stochastic process, which is no longer Markov : 
the distribution of $(R_{i+1},V_{i+1},F_{i+1})$ given the past 
now depends on the whole past and not only on $(R_{i},V_{i},F_{i})$. 
This appears explicitely later on. This is similar to problems 
encountered when studying 
Hidden Markov Models (\citealp{Genon:2003b, Genon:2006, Cappe:2005}). 
For a first approach, we restrict our attention to a very 
informative example, the case of Poisson  distributions, which leads 
to explicit  computations.

\subsection{Probabilistic properties in the Poisson case}
Let us specify all the distributions appearing in the populations dynamics. 
\begin{assumption}
	\label{GI:P} The offspring distribution $G(.)$ is 
Poisson ${\cal P}(m)$, 
the immigration distribution $\mu(.)$ is Poisson ${\cal P}(u)$. 
\end{assumption}
\begin{assumption}
	\label{ST:P} The variables $S_0$ and $T_0$ are independent and 
distributed according 
to Poisson distributions: $S_0 \sim {\cal P}(\sigma)$ 
and $T_0 \sim {\cal P}(\tau)$. 
\end{assumption}
For Poisson distributions, we denote 
${\cal P}(\lambda)(k)=\mathbb{P}(X=k)$.
Recall a property of Multinomial and Poisson distributions. 
\begin{lemma}\label{lem:MP} 
Assume that N is a random variable distributed according 
to a Poisson distribution ${\cal P}(\lambda)$ and that 
$X=(X_{1},X_{2},\ldots,X_{l})$ 
is a l-dimensional random variable such that the conditional 
distribution of $X$ given $N$ is
a Multinomial distribution ${\cal M}(N; a_{1},\ldots,a_{l})$ 
with $\sum _{i=1}^{l} a_{i}=1$. 
Then, the random variables $\{X_{i}, i=1,\ldots, l\}$ are independent 
and verify  $X_{i}\sim  {\cal P}( a_{i}\lambda)$. 
\end{lemma}

First consider one population and omit the index $k$ in what follows. Set 
\begin{equation}\label{def:YGi} 
Y_i=(R_i,V_i, F_i) \;\;\mbox{ and }
 \;\;{\cal G}_{i}= \sigma(Y_j,\; j=0,\dots,i).	
\end{equation}
Clearly, ${\cal G}_{i}$ is the information available up to time $i$. 
To state the main result of this section, define the three sequences
of ${\cal G}_{i-1}$ 
measurable random variables: 
\begin{equation}\label{def:Gamma0} 
\Gamma_0=\sigma\;;\;\Gamma '_0=\tau\; ;\;	
\end{equation}
\begin{equation}
	\label{def:Gammai} \mbox{ for } i\geq 1 ,\;\;\Gamma_i = a \Gamma_{i-1} +a' \Gamma '_{i-1}\;;\; \; \Gamma'_i= m F_{i-1}+u . 
\end{equation}
\begin{equation}
	\label{def:Lambdai} \mbox{ for } i\geq 0,\;\; \Lambda_i =
 b \Gamma_{i} +b' \Gamma '_{i}. 
\end{equation}
Then, the following holds :  
\begin{theorem}\label{theo:cond} 
Under Assumptions (A\ref{G:iid})-(A\ref{ST:P}), 
the initial distribution $\tilde{\pi}_0(y)$ of $(Y_i)$, 
and the conditional distribution 
$ {\cal L}(Y_{i+1}/{\cal G}_{i})$ satisfy,
 using (\ref{def:YGi}) and definitions 
(\ref{def:p4p5}),(\ref{def:Lambdai}), for $y=(r,v,f)
\in \mathbb{N}^3$
\begin{align}\label{eq:RVFi} 
\mathbb{P}(Y_{0}=(r_0,v_0,f_0)) =\tilde{\pi}_0(y_0) &=
{\cal P}(\Lambda_0)(r_0) p_4(v_0/r_0) p_5(f_0/v_0),\\ 
\mathbb{P}(Y_{i+1} =(r,v,f)/{\cal G}_{i}) &= 
{\cal P}(\Lambda_{i+1}) (r) p_4(v/r) p_5(f/v). 
\end{align}
\end{theorem}
The explicit dependence of $R_{i}$ on
the whole past $(Y_0,\dots,Y_{i-1})$ appears more 
simply with the following expression for 
$\Lambda_i$,
\begin{align}\label{eq:Lambdai}
\Lambda_0 & = b\sigma+b'\tau = c_0(\theta)\;;\;\\
\Lambda_1 & = b'm F_0+ c_1(\theta)  \; \mbox{ with } c_1(\theta)= ab\sigma+a'b\tau+b'u,\\ 
\Lambda_i  & = b'm F_{i-1} +
 a'bm (F_{i-2}+a F_{i-3}+\dots+ a^{i-2}F_0)
+ c_i(\theta), \mbox{ with}\\
& c_i(\theta)  =  a^{i}b\sigma+ a^{i-1}a'b\tau +
a'bu\;\frac{1-a^{i-1}}{1-a}+ b'u \;\mbox{ for  } i\geq 2.
\end{align}
 
The  result of Theorem \ref{theo:cond} is a consequence 
of the proposition stated below. 
\begin{proposition}\label{prop:cond} Under Assumptions
        (A\ref{G:iid})-(A\ref{ST:P}), 
the random variables $S_{i+1}$, $T_{i+1} $ are conditionally 
independent given ${\cal G}_{i}$, and their conditional distributions 
satisfy, using the random variables $\Gamma_i$ and $\Gamma'_i$ 
defined in (\ref{def:Gamma0}), 
(\ref{def:Gammai}), 
\begin{equation}\label{eq:condSiTi}
{\cal L}(S_{i+1}/{\cal G}_{i}) \sim {\cal P}(\Gamma _{i+1})\; 
\mbox{and }\; {\cal L}(T_{i+1}/{\cal G}_{i}) \sim {\cal P}(\Gamma' _{i+1} ). 
\end{equation}
\end{proposition}
Let us prove Theorem \ref{theo:cond} assuming Proposition \ref{prop:cond}.
We just have to check the expression of the conditional distribution 
of $R_{i+1}$. 
By (\ref{def:S'R'}), we have $R_{i+1}=R'_{i+1} 
+R''_{i+1}$. Using Proposition \ref{prop:Markov} and (\ref{def:p2p3}), 
the distribution of $R_{i+1}$ conditionnally on
$(S_{i+1},T_{i+1})$ is equal to $p_{3}(r/S_{i+1},T_{i+1}) $. 
Applying Proposition \ref{prop:cond}, $S_{i+1}$ and $T_{i+1}$ are 
conditionally independent given ${\cal G}_{i}$ and distributed 
according to two independent 
Poisson distributions. Hence, another application of Lemma
\ref{lem:MP} yields that the conditional distribution of $R_{i+1}$
given ${\cal G}_{i}$ is ${\cal P}(b\Gamma_{i+1}+ b' \Gamma'_{i+1})=
{\cal P}(\Lambda_{i+1})$, which is (\ref{def:Lambdai}).

The proof of Proposition \ref{prop:cond} is given in Appendix 
\ref{appen:cond}.

\subsection{Likelihood of the incomplete observations}

The inference is now based on the observations of $(Y_i)$ recorded up to time
$n$ for $K$ independent populations. Denote by
$Y_{i}^{k}=(R_{i}^{k},V_{i}^{k}, F_{i}^{k})$ the process
describing its dynamics in population $k$ and  
$y_{i}^{k}=( r_{i}^{k},v_{i}^{k}, f_{i}^{k})$ the observations at time
$i$. We set,
\begin{equation}\label{def:y} 
 Y_{0:n}^k=(Y_0^k, \dots,Y_n^k),\;\; \mbox {and }\;\; Y_{0:n}(K)=(Y_{0:n}^1,\dots,Y_{0:n}^K).    
 \end{equation}
Observations up to time $n$ are denoted 
\begin{equation}\label{def:Otilde} 
\tilde{O}_{0:n}^{k}=(y_{0}^{k},
        y_{1}^{k},\dots, y_{n}^{k})\;\; 
\mbox{and }
\;\tilde{O}_{0:n}(K) = (\tilde{O}_{0:n}^{1},\dots,\tilde{O}_{0:n}^{K}). 
\end{equation}
Let us first compute the likelihood for one population, population  $k$.
Successive conditionnings yield 
\begin{equation} \label{def:Ltildek} 
L(\theta ;\tilde{O}_{0:n}^k)  
	=P_{\theta}(Y_{0}^k=y_{0}^k)\prod_{i=1}^{n} P_{\theta}(Y_{i}^k=
        y_{i}^k/Y_{0:i-1}^k=
 y_{0:i-1}^k). 
\end{equation}
Contrary to the previous section, each term of this product depends on
$i$ and on the observations up to time $i-1$. Theorem \ref{theo:cond}
 gives the expression of 
these conditional probabilities. 

Since the random variables $\Lambda_i$ now depend on
 $\theta$ and on the past, we define 
$\Lambda _{i}(\theta)= \Lambda_i(\theta;
 Y_{0:i-1}) $, and for population $k$, 
\begin{equation}
	\label{def:Lambdaik} \Lambda _{i}^{k}(\theta)=
 \Lambda_i(\theta; Y_{0:i-1}^k)\;,\; \;
\lambda _{i}^{k}(\theta)
=\Lambda_{i} (\theta;y_{0:i-1}^k). 
\end{equation}
\begin{equation}
	\label{def:ll} P_{\theta}(Y_{i}^k= y_{i}^k/
y_{0:i-1}^k) 
= {\cal P}(\lambda _{i}^{k}(\theta))(r_{i}^k)\; 
p_3(c;v_{i}^k/r_{i}^k)\;p_4(d;f_{i}^k/v_{i}^k). 
\end{equation}
Joining the observations in the $K$ populations, the likelihood 
writes, using notations (\ref{def:y}), (\ref{def:Otilde}), (\ref{def:ll}),
\begin{equation}\label{def:Ltilde} 
L(\theta ;\tilde{O}_{0:n}(K))= \prod_{k=1}^{K} L(\theta ;\tilde{O}_{0:n}^k).
\end{equation}
The log-likelihood splits into three terms, 
\begin{equation}\label{def:ltilde} 
	\tilde{l}(\theta,\tilde{O}_{0:n}(K)) = \log L(\theta ;\tilde{O}_{0:n}(K) )=
 \sum_{i=1}^3\tilde{l}_K^{\;i}(\theta,\tilde{O}_{0:n}(K)),
\end{equation}
where, using Assumptions (A\ref{GI:P}), (A\ref{ST:P}) and 
Theorem \ref{theo:cond}, 
\begin{equation}
	\label{def:ltilde1} 
\tilde{l}_K^{\;1}(\theta)= \tilde{l}_K^{\;1}(\theta,\tilde{O}_{0:n}(K))= 
	\sum_{k=1}^K \sum_{i=0}^n \log {\cal P}(\lambda _{i}^{k})(r_{i}^k), 
\end{equation}
\begin{equation}
	\label{def:ltilde2} 
\tilde{l}_K^{\;2}(\theta,\tilde{O}_{0:n}(K))=
        l_K^2(c,O_{0:n})\;;
\; \tilde{l}_K^{\;3}(\theta,\tilde{O}_{0:n}(K)) =l_K^ 3(d,O_{0:n}). 
\end{equation}
Hence, estimating parameters $c,d$ is exactly the same as 
in the previous section; 
their inference is omitted in the sequel.

\subsection{Parametric inference}
It remains to study the estimation of 
the parameter   $\theta= (\sigma,\tau,a,b,a',b',u,m)$. As before,
let  $\theta_ 0$ be
true value of the parameter. 

We first have to investigate which parameters are identifiable
when only these incomplete observations are available.
By identifiability, we  mean here identifiability of a  statistical 
model ${\cal M}= (\mathbb{P}_\theta, \theta \in \Theta )$ :
$$\forall \theta,\;\theta' \in \Theta,
\{ \mathbb{P}_{\theta} = \mathbb{P}_{\theta'}\} \Rightarrow \{ \theta
=\theta'\}.$$

Recall that $n$ denotes the time index of a plant lifecycle  
and that we consider populations recorded up to time $n$ ($n=0$ corresponding 
here to the observation of one complete cycle). Using now the definitions of the terms  
$c_i(\theta)$ given in (\ref{eq:Lambdai}), the following holds :
\begin{theorem}\label{theo:identifiability}
 Assume  (A\ref{G:iid}), (A\ref{SiTi}), (A\ref{a:tind}), (A\ref{ass:param}), 
(A\ref{GI:P}), (A\ref{ST:P}). Then,
\begin{enumerate}
\item  if $n=0$, only $ c_0(\theta) =b\sigma+b'\tau$ is identifiable;
\item if $n=1$, $(b'm, c_0(\theta), c_1(\theta))$ is identifiable; 
\item if $n=2$, $(\frac{a'b}{b'},b'm, c_0(\theta), c_1(\theta),
c_2(\theta))$,  is identifiable;
\item if $n\geq 3$, and $a\neq \frac{a'b}{b'}$,
then $\phi= (a, \frac{a'b}{b'},b'm, b'u, b\sigma, b'\tau)$ is identifiable; \\
if $n \geq 3$ and $a=\frac{a'b}{b'}$, then
$(a=\frac{a'b}{b'},b'm, b'u, b\sigma+b'\tau)$.
\end{enumerate}
\end{theorem}
Note that identifiability of additional parameters cannot be gained
increasing $n$ beyond 3. Larger values of $n$  result in improving the
asymptotic variance for the estimation of $\phi$.

\begin{remark}\label{rem:i} 
Stating the above theorem for the first values 
of $n$ is unusual. However, in the field survey of feral 
oilseed rape populations, observations unfortunately had been 
collected up to $n=2$, leading to the 
unability of estimating $a$, the annual survival rate in the seed
bank, which is a parameter of much concern in Ecology. 
\end{remark}

\bigskip

Assume now that $n\geq 3$, $a\neq \frac{a'b}{b'}$ and let us study the inference of the
identifiable parameters. Let us denote by $\phi= (a, \frac{a'b}{b'},b'm, b'u, b\sigma, b'\tau) $ 
(resp. $\phi_0$)
an arbitrary (resp. the true value) of the paramete, by $\Phi$ the parameter set and asssume

\begin{assumption}\label{ass:Phi}
$\Phi$ is a compact set of $\mathbb{R_{+}^{*}}^5 \times (0,1)$; and 	
$\phi_0 \in \overset{\circ}{\Phi}. $
\end{assumption}
\noindent
Using (\ref{def:ltilde1}), define the maximum likelihood estimator
$\hat{\phi}_K $ as a solution of
\begin{equation}\label{def:phiK}  
\tilde{l}_K^{\;1}(\hat{\phi}_K) 
= \sup \{\tilde{l}_K^{\;1}(\phi)\;;\phi \in \Phi\}.
\end{equation}
\noindent
Under Assumption (A\ref{ass:Phi}), the function $\phi \rightarrow  \sum _0 ^n (\Lambda_i(\phi;Y_{0:i-1})-\log \Lambda_i(\phi,Y_{0:i-1}))$
is a.s. $\mathbb{P}_{\phi}$ twice differentiable on $\Phi$, and we can define, 
for $\phi=(\phi_1,\dots,\phi_6)$, the $6\times 6$ matrix,  
$$
I(\phi)_{p,q}= \sum_{i=0}^{n}\mathbb{E} _{\theta_0}(
\frac{1}{\Lambda_i(\phi)}\frac{\partial\Lambda_i(\phi)}{\partial\phi_p}
\frac{\partial\Lambda_i(\phi)}{\partial\phi_q}) 
\; \; \mbox{with  }  1 \leq p,q \leq 6.
$$
 
\begin{proposition}\label{prop:statphi}
Assume (A\ref{G:iid})-(A\ref{a:tind}), (A\ref{GI:P}), (A\ref{ST:P}) 
and (A\ref{ass:Phi}). 
Then $\hat{\phi}_K $ is
strongly consistent.
If moreover the matrix $I(\phi_0)$ is invertible, then
$$ 
  \sqrt{K}(\hat{\phi}_K -\phi_0)\overset{{\cal D}} {\rightarrow}  
{\cal N}(0,I(\phi_0)^{-1}) \mbox{ under } \mathbb{P}_{\phi_0}
\mbox{ as }  K\rightarrow  +\infty .
$$
\end{proposition}

\begin{remark}\label{rem:statIphi}
The matrix $I(\phi_0)$ that appears in the asymptotic
variance of $ \hat{\phi}_K$ can be estimated, using the explicit expressions 
for the derivatives of $\Lambda_i(\phi)$, by the empirical 
estimate for $1\leq p,q \leq 6 $,
$$
\hat{I}_{p,q}=\frac{1}{K} \sum_{k=1}^{K} \sum_{i=0}^{n}
\frac{1}{\lambda_i^k(\hat{\phi}_K)}\frac{\partial\lambda_i^k(\hat{\phi}_K)}
{\partial\phi_p}
\frac{\partial\lambda_i^k(\hat{\phi}_K)}{\partial\phi_q}.
$$
\end{remark}
The proofs of Theorem 
\ref{theo:identifiability} and Proposition \ref{prop:statphi}
are  given in Appendix \ref{appen:identifiability}.\ref{appen:statphi}.

\section{Simulation study}\label{sect:simul}
In the case of incomplete observations, the
estimators strongly rely on the assumption that both the offspring
and the immigration distributions are Poisson distributions. We
investigate in this section how the estimation behaves when these
distributions are no longer Poisson.

\subsection{Methods}\label{sub:simulmethods}
We considered in these simulations that 
the offspring distribution $G(.)$ or the immigration 
distribution $\mu(.)$ are  more dispersed than Poisson distributions. 
Two series of simulations were performed: the first one concerns
deviations to a Poisson distribution of the offspring $G(.)$, and the other one for the immigration $\mu(.)$.
For this, we used  Negative Binomial distributions for $G(.)$ (resp. $\mu(.)$)
with mean $m$ (resp.
$u$) and increasing 
values of the variance variance $\delta^2$ (resp. $\rho^2$ ); the 
(variance /mean) ratio ranged from 2 to 1000. For each 
given set of parameters, we performed 
$M = 100$ repetitions including $K =300$ populations. These populations 
dynamics  were run during  $n= 4$ years to get the observations for 
rosettes $R_0$ to $R_4$ and mature plants ($F_0$ to $F_3$), 
using biologically plausible values for demographic 
parameters. Indeed, for $a,b,a',b'$, 
we used the parameters given in 
\citealp{Claessen:2005a, Claessen:2005b} : 
\begin{equation}\label{val:ab}
a =0.15\;,\; a'=0.006\;, \;b=0.5\;, \; b'=0.5 .
\end{equation} 
We also had to fix some values for $c,d, m=  \mathbb{E}(G), u =\mathbb{E}(\mu)$.
We used values estimated in \citealp{Garnier:2008} :
\begin{equation}\label{val:cdmu}
c=0.21\;,\;d=0.01\;,\; m=13\; , \;u=80.
\end{equation}
The value $m=13$  corresponds to the mean fecundity of plants  mown twice.
(\citealp{Colbach:2001a}). 
The value $u=80$ corresponds to average immigration when there is 
no cultivated field
in the neighbourhood (\citealp{Garnier:2008}).
Finally, for the initial distributions of $S_0, T_0$, 
we assumed 
\begin{equation}\label{val:ST0}
S_0 \sim {\cal P}(\sigma), T_0 \sim  {\cal P}(\tau)
\mbox{ with } \sigma=\tau = 50.
\end{equation}
For each value of (variance/mean) ratio, the inference of the 
identifiable parameters (see Theorem \ref{theo:identifiability})
in the case $n\geq 3$. was computed on each of the $M=100$
repetitions of the $K=300$ populations  
trajectories. Mean and standard deviation of these estimates were 
then empirically estimated from the $M=100$ values obtained.
Simulations and statistical analyses were performed using R-8-0 (\citealp{Rteam:2005}).

\subsection{Results}\label{sub:simulresults}
Some technical difficulties occured when we tried to estimate jointly
the six identifiable parameters
$(a,\frac{a'b}{b'},b'm,b'u, b\sigma,b'\tau)$: the algorithm we 
used often did not converge because of the non-linearity in the statistical model.
Since we were mainly interested in the estimations for $G(.)$ and $\mu(.)$,
we assumed that the quantities $a$ and $\frac{a'b}{b'}$ were known.
With this simplification, we just had to deal with a
linear statistical model; we restricted our 
attention to the estimation of the parameters 
$(b'm, b'u, b\sigma, b'\tau)$.

The results are given in the two tables below.

\begin{table}[htp]\label{table:G}
\centering
\begin{tabular}{|c|cc|cc|cc|cc|}

\hline
& \multicolumn{2}{|c}{$b'm=6.5$} & \multicolumn{2}{|c}{$b'u=40$} & \multicolumn{2}{|c}{$b'\sigma=25$} & \multicolumn{2}{|c|}{$b'\tau=25$} \\

$\delta^2/m$ & est. & sd & est. & sd & est. & sd & est. & sd \\
\hline
2 & 6.44 & 0.77 & 40.02 & 0.21 & 25.19 & 3.28 & 24.88 & 3.27 \\

5 & 6.46 & 0.61 & 40.05 & 0.24 & 24.46 & 3.34 & 25.52 & 3.36 \\

10 & 6.65 & 0.84 & 39.96 & 0.23 & 25.18 & 2.93 & 24.84 & 2.95 \\

50 & 6.78 & 1.40 & 39.97 & 0.27 & 25.10 & 3.73 & 24.88 & 3.75 \\

100 & 6.56 & 1.89 & 39.99 & 0.25 & 25.24 & 3.63 & 24.73 & 3.64 \\

500 & 6.26 & 3.30 & 40.01 & 0.32 & 24.45 & 6.66 & 25.50 & 6.69 \\

1000 & 6.41 & 5.41 & 40.003 & 0.38 & 25.54 & 7.75 & 24.37 & 7.78 \\

\hline
\end{tabular}

\caption{Mean (est.) and standard deviation (sd) of estimators when the offspring $G(.)$ 
is a binomial distribution 
 with mean $ m$ and variance $\delta ^2$ instead of a ${\cal P}(m)$ distribution.
Immigration $\mu \sim {\cal P}(u)$;
 $ a=0.16, a'=0.006, b=b'=0.5,  m=13, u= 80, \sigma=\tau=50$.}
\end{table}

\begin{table}[htp]\label{table:mu}
\centering
\begin{tabular}{|c|cc|cc|cc|cc|}
\hline
Value & \multicolumn{2}{c}{$b'm=6.5$} & \multicolumn{2}{|c}{$b'u=40$} &
 \multicolumn{2}{|c}{$b'\sigma=25$} & \multicolumn{2}{|c|}{$b'\tau=25$} \\

$\rho^2/u$ & est. & sd & est. & sd & est. & sd & est. & sd \\
\hline
2 & 6.51 & 0.77 & 40.01 & 0.24 & 24.61 & 3.81 & 25.36 & 3.76 \\

5 & 6.45 & 0.98 & 40.01 & 0.37 & 24.33 & 5.19 & 25.67 & 5.28 \\

10 & 6.61 & 1.47 & 40.03 & 0.52 & 24.78 & 7.25 & 25.17 & 7.25 \\

50 & 6.59 & 2.63 & 39.93 & 0.87 & 25.60 & 14.89 & 24.46 & 14.93 \\

100 & 7.42 & 3.43 & 39.95 & 1.38 & 28.60 & 22.42 & 21.39 & 22.40 \\

500 & 7.61 & 7.26 & 40.05 & 3.13 & 21.96 & 39.31 & 28.06 & 39.31 \\

1000 & 6.48 & 7.63 & 39.62 & 4.23 & 21.77 & 64.75 & 28.23 & 64.76 \\
\hline
\end{tabular}

\caption{Mean (est.) and standard deviation (sd) of estimators 
when the immigration $\mu(.)$ is
a negative binomial distribution with mean $u$ and variance
$\rho^2$ instead of a ${\cal P}(u)$ ditribution.
Offspring $G(.) \sim {\cal P}(m)$; $a=0.16, a'=0.006, b=b'=0.5,  m=13, u= 80,
\sigma=\tau=50$.}
\label{table:immigration}
\end{table}

Concerning deviations of $G(.)$ from the distribution ${\cal P}(m)$,
 we obtained 
that the estimation procedure performed very well, even for large
deviations from the Poisson case : the bias remained less than 5\% 
for all the four estimated quantities with  a variance/ratio up to 1000. 
(see Table \ref{table:G})
When immigration was assumed to follow a Negative Binomial
distribution, 
the estimation procedure performed quite well for values of 
(variance/mean) ratios up to 50 : the biases remained less than 10\% 
for all four identifiable quantities. For larger values of 
variance/mean ratio, the bias could 17\% of parameter value approximately.

\bigskip

\appendix
\section{Appendix section}\label{app}

\subsection{Proof of Proposition \ref{prop:Markov} }
\label{appen:Markov}
Consider first the initial distribution $\pi _{0} (x) $.
Successive conditionings yield
$$\pi_0 (x) = \mathbb{P}(S_0=s, T_0=t)\mathbb{P}(R_0=r/s,t)
\mathbb{P}(V_0=v/s,t,r)\mathbb{P}(F_0= f/s,t,r,v).$$
The first distribution is $\nu(s,t)$. 
Using  definitions  (\ref{def:cd}),
 (\ref{def:p4p5}), the last two  conditional distributions 
are equal to $ p_{4}(v/r)$ and $p_{5}(f/v)$.
The remaining distribution in $\pi_0 (x) $ is 
${\cal L}(R_0/S_0, T_0)$. Using (A\ref{SiTi}) and (\ref{def:S'R'}),
$R_0 = R'_0 + R_{0}''$, where $R'_0$ and  $R_{0}''$ are independent.
Now, the distribution of 
$(S'_1, R'_0)$ (resp. $(S_{1}'',R_{0}'')$) is  the Multinomial distribution
${\cal M}(S_0; a,b)$  (resp. $  {\cal M}(T_0; a',b')$), leading to  
Binomial distributions 
${\cal B}(S_0; b)$ for the marginal distribution of $R'_{0}$
(resp. ${\cal B}(T_0; b')$ for  $R''_0$ ).
By  (A\ref{SiTi}), these two distributions are independent conditionally on 
$(S_0,T_0)$ which  
yields  (\ref{def:p2p3}) and the expression of $\pi_0 (x) $.
Let us now study two successive generations. 
Using notations (\ref{def:X}), the conditional
distribution of $X_{i+1}$ given
${\cal F} _i$ can be expressed for $x= (s,t,r,v,f),x'=(s',t',r',v',f')$, 
\begin{multline*}
\mathbb{P}(X_{i+1}=x'/{\cal F}_{i}) = \mathbb{P}(S_{i+1}=s'/{\cal F}_{i})
\times \mathbb{P}(T_{i+1}=t'/s'; {\cal F}_{i})\times\\
\mathbb{P}(R_{i+1}=r'/s',t';{\cal F}_{i})
\times
\mathbb{P} (V_{i+1}=v'/s',t',r';{\cal F}_{i})\times
\mathbb{P}(F_{i+1}=f'/s',t',r',v';{\cal F}_{i}). 
\end{multline*}
The last three conditional distributions have already been computed
for getting $\pi_0(x)$. They 
are respectively equal to $p_3(r'/s',t')$, $p_4(v'/r')$ and  
$p_5(f'/v')$ defined in (\ref{def:p2p3}), (\ref{def:p4p5}). Let us compute
$\mathbb{P}(T_{i+1}=t'/s'; {\cal F}_{i})$.
Seeds on the ground at cycle $(i+1)$ come from two sources, offspring
of mature plants $F_{i}$ and  seed immigration during cycle $i+1$,
$T_{i+1} =  T'_{i+1} + I_{i+1}$. 
Using (A\ref{G:iid}), the distribution of $T'_{i+1}$ is $G ^{\star F_{i}}$.
By (A\ref{a:tind}), $I_{i+1}$ is independent of 
 $S_{i+1},T'_{i+1}$ and ${\cal F}_{i}$, so that 
 $ \mathbb{P}(T_{i+1}=t'/s'; {\cal F}_{i}) = (G ^{\star f} 
\star \mu) (t')= p_2(t'/f)$.
The last distribution is $\mathbb{P}(S_{i+1}=s'/{\cal F}_{i})$. 
Using (A\ref{SiTi}) and (\ref{def:S'R'}) yields that
 $S_{i+1} = S'_{i+1} + S_{i+1}'' $
 where  
conditionally on  $(S_{i}, T_{i})$, $(S'_{i+1}, R'_i)$ 
and  $(S''_{i+1}, R_{i}'')$ are distributed according to two
independent Multinomial distributions 
${\cal M}(S_i; a,b)$ and ${\cal M}(T_i; a',b')$. Hence, the
marginal distribution of $S_{i+1}$ is,
$$
 \mathbb{P}(S_{i+1}=s'/ {\cal F}_i) = \frac{({\cal M}(S_{i}; a,b)
\star{\cal M}(T_{i}; a',b'))(s', r)}{(B(S_{i},b)
\star B(T_{i},b'))(r) }= p_{1}(s'/s,t,r).
$$
Joining these results yields that  
$(X_i)$ is a time homogeneous 
Markov chain with state space $\mathbb{N}^5$. 

\subsection{Proof of Proposition \ref{prop:mle}}\label{appen:licomp}
We use in the sequel the Kullback-Leibler divergence 
${\cal K}(P,Q)$ of
distribution $Q$  w.r.t. $P$. 
Recall its  definition,
\begin{align}\label{def:Kullback}
{\cal K}(P,Q) &= -\int \; \log \frac{dQ}{dP}\;dP =-\mathbb{E}_{\mathbb{P}}(\frac{dQ}{dP}) 
\;\mbox{if } Q \ll P\; ; \;\\
{\cal K}(P,Q) &=+ \infty \;\;\mbox{ otherwise}.
\end{align}
This quantity is non-negative and equal to 0 if and only if $Q=P\;
\;P$ a.s.

Let us first consider $l_K^2( c)$ 
and $l_K^3(d)$ defined in (\ref{eq:l2}), (\ref{eq:l3}).
The maximum likelihood estimators are 
\begin{equation}\label{def:cdK}
\hat{c}_K=\frac{\sum _{k=1}^{K} 
 \sum _{i=0}^{n} \; v_{i}^{k}}{\sum _{k=1}^{K} 
 \sum _{i=0}^{n} \; r_{i}^{k}}\;\;;\;\; 
\hat{d}_K=\frac{\sum _{k=1}^{K} 
 \sum _{i=0}^{n} \; f_{i}^{k}}{\sum _{k=1}^{K} 
 \sum _{i=0}^{n} \; v_{i}^{k}}.
\end{equation}
Since the $K$ populations are independent,
applying the strong law of large numbers yields the strong consistency 
of $\hat{c}_K$ and $\hat{d}_K$. The random variables
$Z_i^k= \sum_{i=0}^n (V_i^k-c_0 R_i^k)$ are i.i.d. centered with 
variance $c_0(1-c_0)\mathbb{E}_{\theta_0} (\sum_{i=0}^nR_i)$, so that the  
Central Limit Theorem yields that,

\begin{equation}\label{eq:cK}
\mbox{as }  K \rightarrow +\infty \;,\;	\sqrt K (\hat{c}_K-c_0) \overset{{\cal D}}{\rightarrow }
{\cal N}(0, \frac{c_0(1-c_0)}{\mathbb{E}_{\theta _0}(\sum _{i=0}^{n} R_{i})})
\;\;\mbox{under  } \mathbb{P}_{\theta_0}.
\end{equation}
The proof concerning $\hat{d}_k$ is similar : $ \hat{d}_K
\rightarrow d_0 \; \mathbb{P}_{\theta_0}$ a.s. and
\begin{equation}\label{eq:dK}
\sqrt K (\hat{d}_K-d_0) \overset{{\cal D}}{\rightarrow }
{\cal N}(0, \frac{d_0(1-d_0)}{\mathbb{E}_{\theta _0}(\sum _{i=0}^{n} V_{i})}).
\end{equation}

Consider now the estimation of $(a,b,a',b')$. Let us 
first check identifiability. Applying the 
strong law of large numbers to $ l_K^6 (\theta;{\cal O}_{0:n})$ defined
in (\ref{eq:l6}), we get that, under  $\mathbb{P}_{\theta_0}$,
as $K \rightarrow +\infty$,
\begin{equation*}
\frac{1}{K} l_K^6 (\theta; X_{0:n}(K)) 
\rightarrow \mathbb{E}_{\theta_0}\sum _{i=0}^{n} 
 \log({\cal M}(S_{i}; a,b) \star {\cal M}(T_{i}; a',b')) (S_{i+1},R_i).
\end{equation*}
We can express the limit above using the Kullback-Leibler divergence
defined in (\ref{def:Kullback}),
$$ 
C(\theta_0) -  \mathbb{E}_{\theta_0} 
 \sum _{i=0}^{n} {\cal K}({\cal M}(S_{i}; a_0,b_0) 
\star {\cal M}(T_{i}; a_0',b'_0),{\cal M}(S_{i}; a,b) 
\star {\cal M}(T_{i}; a',b')),
$$
where $C(\theta_0)$ is a constant depending only on $\theta_0$,
and $ {\cal K}(P,Q)$ is the Kullback- Leibler divergence of the two random
probability distributions.
Each term of the sum
above is  non positive  and equal to 0 if and only if  the   
two distributions are identical a.s. under $\mathbb{P}_{\theta_0}$.
It is easy to check, using the first and second moments of these 
two distributions, that
this implies $(a,b,a',b')=(a_0,b_0,a'_0,b'_0)$. Hence,$(a_0,b_0,a'_0,b'_0)$    
is a strict maximum of the limit above,
which leads to the identifiability of these parameters. 

Consider now the CLS estimators $(\hat{a}_K,\hat{a}'_K) $ which minimize 
$J_K^1(a,a')$
defined in (\ref{def:J1J2}).
By the strong law of large numbers, under $\mathbb{P}_{\theta_0}$, as
$K \rightarrow\infty$, 
\begin{equation}\label{eq:limJ}
\frac{1}{K} J_K^1(a,a') 
\rightarrow  
\sum _{i=0}^{n}\mathbb{E}_{\theta_0}(\{(a_0-a)S_i+ (a'_0-a')T_i\}^2)
 + A(\theta_0),
\end{equation}
where $A(\theta_0)=a_0(1-a_0)E_{\theta_0}(\sum _{i=0}^{n}S_i) +
a'_0(1-a'_0)\mathbb{E}_{\theta_0}(\sum _{i=0}^{n}\mathbb{E}_{\theta_0}T_i)$ 
only depends on $\theta_0$.
Since $S_i$ and $T_i$ are non negative random variables,
this  limit  possesses a strict minimum at $(a,a') =(a_0,a'_0)$.

Denote by $^\tau M$ the transposition of a matrix $M$ and set 
$Z$  the $(n\times 2)$ matrix with rows equal to
$(S_i,T_i)$,
$^\tau S$ the vector $(S_0,\dots,S_n)$, $^\tau \tilde{S}$ the
vector $(S_1,\dots,S_{n+1})$,
and $Z^k,
S^k,\tilde{S}^k,T^k$ their values for population $k$. 
Then $ J_K^1(a,a')$ writes, 
$$
\frac{1}{K} J_K^1(a,a')=\frac{1}{K} \sum_{i=1,\dots,K}\;
 ^{\tau}\biggl(\tilde{S}^k- Z^k\begin{pmatrix}
a\\a'
\end{pmatrix}\biggr)\biggl(\tilde{S}^k- Z^k\begin{pmatrix}
a\\a'
\end{pmatrix}\biggr)\;
$$
\begin{equation}\label{def:ak}
 \mbox{ so that  }\begin{pmatrix}
	\hat{a}_K\\\hat{a}'_K 
\end{pmatrix} =\biggl( \sum _{k=1}^K\; ^\tau Z^k \; Z^k \biggr)^{-1}
 \biggl( \sum _{k=1}^K \; ^\tau Z^k \tilde{S^k} \biggr).
\end{equation}
As functions of $(a,a')$, 
$\frac{1}{K}J_K^1(a,a')$ and its limit defined in (\ref{eq:limJ}) 
are twice continuously differentiable 
a.s.. The parameter
set $\Theta$ is compact, so  we just have to
control the continuity modulus 
of the process $\frac{1}{K} J_K^1(a,a')$. It is defined, for $\eta >0$, by
$$
w(K,\eta)= sup\{\frac{1}{K}  \vert J_K^1(a_1,a'_1)-J_K^1(a_2,a'_2)\vert\;;\; 
 \Vert (a_1,a'_1)-(a_2,a'_2) \Vert < \eta\}.
$$
By the Cauchy-Schwarz inequality, $w(K,\eta)$ is bounded by 
$$
\{\frac{1}{K} \sum _{k,i}(2S_{i+1}^k- (a_1 +a_2) S_{i}^k
-(a'_1 +a'_2) T_{i}^k)^2 \}^{1/2} \;\{ \frac{1}{K}\sum _{k,i}
(( a_1 -a_2) S_{i}^k +(a'_1-a'_2)T_{i}^k)^2 \} ^{1/2}.
$$
The first term is a random variable converging
$\mathbb{P}_{\theta_0}$-a.s. to a deterministic 
positive limit,
and is thus bounded in probability. The second term is bounded
 by the r.v. 
$ 2 \eta (\frac{1}{K}\sum _{k=1}^{K}\sum _{i=0}^{n}(S_{i}^k
 +T_{i}^k))^{1/2} \rightarrow  2 \eta \;(E_{\theta_0}\sum _{i=0}^{n}(S_{i}+T_i))^{1/2}$ 
$\mathbb{P}_{\theta_0}$ a.s.Ò
Hence, as $K\rightarrow +\infty$, limsup $ w(K,\eta)\leq \phi (\eta)$, where  
$\phi(\eta) \rightarrow 0$ as $\eta \rightarrow 0$.
This ensures the consistency of
$(\hat{a}_K,\hat {a}'_K) $ defined in (\ref{def:ak}) 
(\citealp{Dacunha:1993,VanderVaart:1998}).	

Consider now the asymptotic normality. 
The function 
$(a,a')\rightarrow J_K^1(a,a')$ being $C^2$ a.s.,
the gradient of $J_K^1(a,a')$ is
$$DJ_K^1(a,a') 
= - \sum_{k=1}^K 
 {^ \tau {Z^k}} \biggl(\tilde{S}^k- Z^k\begin{pmatrix}
a\\a'
\end{pmatrix}\biggr).
$$
The 2x2 matrix containing the second derivatives of
$J_K^1(a,a')$ is 
$$\nabla J_K^1(a,a') = 
2 \sum_{k=1}^K \;^{\tau}Z^k Z^k.$$
Using now that $(\hat{a}_K,\hat{a}'_K)$ is consistent, a Taylor
expansion of $DJ_K^1$ at $(a_0,a'_0)$ yields
\begin{equation}\label{Taylor}
0= \frac{1}{2\sqrt K} DJ_K^1(a_0,a'_0)+ \frac{1}{2\sqrt K}
\nabla J_K^1 \begin{pmatrix}a-a_0\\a'-a'_0 \end{pmatrix}+ o_P(1),
\end{equation}
where ${o}_P(1)$ denotes a remainder term that goes to $0$ 
in $P_{\theta_0}$-probability.
The r.v. vectors $\biggl (\tilde{S} -Z^k\;\begin{pmatrix}
a_0\\a'_0
\end{pmatrix}\biggr)$
are i.i.d. centered, hence   
$\frac{1}{2\sqrt K}DJ_K^1(a_0,a'_0)$
converges in distribution under $\mathbb{P}_{\theta_0}$ to a centered
Normal distribution with covariance matrix 
$E_{\theta_0}(^{\tau}Z V(\theta_0) Z)$ 
where $V(\theta_0) $ is the $(n+1)\times (n+1)$  
diagonal matrix with diagonal elements $ a_0(1-a_0)S_i+
a'_0(1-a'_0)T_i$.
Moreover, by the strong law of large numbers, 
$\frac{1}{2K}\nabla J_K^1$ converges a.s. to 
$E_{\theta_0}(^{\tau}Z Z)$.
By the Cauchy-Schwarz inequality, this matrix is invertible 
and  using (\ref{Taylor}) 
yields 
$$\sqrt K \begin{pmatrix}\hat{a}_K-a_0\\
\hat{a}'_K-a'_0
\end{pmatrix} = \biggl(E_{\theta_0}(^{\tau}Z Z)\biggr)^{-1}
\frac{1}{\sqrt K}\sum_{k=1}^K \; ^{\tau}Z^k (S^k-Z^k
\begin{pmatrix} a_0\\
a'_0 \end{pmatrix}) + o_{P}(1).
$$
Therefore, setting 
$\Sigma_1(\theta_0)=(E_{\theta_0}(^{\tau}Z Z))^{-1} E_{\theta_0}(^{\tau}Z 
V(\theta_0) Z)(E_{\theta_0}(^{\tau}Z Z))^{-1}$,
\begin{equation}\label{lim:Vara}
 \sqrt K \begin{pmatrix}\hat{a}_K-a_0\\
\hat{a}'_K-a'_0
\end{pmatrix} \rightarrow ^{\cal D}
{\cal N}(0, \Sigma_1(\theta_0)).
\end{equation}

Similarly, setting $^{\tau}R$ the vector $(R_0,\dots,R_n)$, define 
\begin{equation}\label{def:bk}
\begin{pmatrix}
	\hat{b}_K\\\hat{b}'_K 
\end{pmatrix} =\biggl( \sum _{k=1}^K\; ^\tau Z^k \; Z^k \biggr)^{-1}
 \biggl( \sum _{k=1}^K \; ^\tau Z^k R^k \biggr).
\end{equation}
The proof concerning 
the estimation of $(b,b')$ is similar; so we get, setting $W(\theta_0)$ 
diagonal matrix with diagonal elements 
$b_0(1-b_0) S_i+b'_0(1-b'_0)T_i$ and 
$ \Sigma_2 (\theta_0)=(E_{\theta_0}(^{\tau}Z Z))^{-1} E_{\theta_0}(^{\tau}Z 
W(\theta_0)Z)(E_{\theta_0}(^{\tau}Z Z))^{-1}$, 
\begin{equation}\label{lim:Varb}
 \sqrt K \begin{pmatrix}\hat{b}_K-b_0\\
\hat{b}'_K-b'_0
\end{pmatrix} \rightarrow ^{\cal D}
{\cal N}(0,  \Sigma_2(\theta_0)).
\end{equation}

Finally, let us study the estimation of $(m,u)$ based 
on the CLS process $J_{K}^4(m,u)$ defined in (\ref{def:J4}). 
Under the assumptions (A\ref{G:iid}), (A\ref{a:tind}), $G(.)$ and  
$\mu(.)$ have finite variances $\delta^2$ and $\rho^2$. Denote by $\delta_0^2 $ and $\rho_0^2 $
the variances associated with $\theta_0$. Then, as $K\rightarrow \infty$,
under  $\mathbb{P}_{\theta_0}$,
$$
\frac{1}{K}J_K^4 (m,u) \rightarrow 
E_{\theta_0}\big(\sum _{i=0}^{n}(\delta_0 ^2 F_i + \rho_0 ^2)\big)+
E_{\theta_0}\big(\sum _{i=0}^{n}[(m_0-m)F_i+(u-u_0)]^2\big).
$$
Clearly, the above functional has a strict minimum at $(m_0,u_0)$, 
leading to the identifiability of $(m,u)$.
 Let $ ^{\tau} F$ denote the vector $(F_0,\dots,F_n)$, $G$ the $(n+1)\times 2$ matrix with rows equal to
$(F_i,1)$, $^{\tau}\tilde{T}$ the vector $(T_1,\dots,T_{n+1})$, $F^k, G^k, \tilde{T}^k$
their values in population $k$.
Then, the conditional least square estimator is
\begin{equation}\label{def:muk}
\begin{pmatrix}
	\hat{m}_K\\\hat{u}_K 
\end{pmatrix} =\biggl( \sum _{k=1}^K\; ^\tau G^k \; G^k \biggr)^{-1}
 \biggl( \sum _{k=1}^K \; ^\tau G^k \tilde{T^k} \biggr).
\end{equation}
Consistency and asymptotic normality are obtained with a proof similar
to the one detailed above. Define $W'(\theta_0)$ the
diagonal matrix with diagonal elements $\delta_0^2 F_i+ \rho_0^2$, and 
$ \Sigma_3(\theta_0)= (E_{\theta_0}(^{\tau}G G))^{-1} E_{\theta_0}(^{\tau}G W'
 G)(E_{\theta_0}(^{\tau}G G))^{-1}$, then
\begin{equation}\label{lim:Varmu}
 \sqrt K \begin{pmatrix}\hat{m}_K-m_0\\
\hat{u}_K-u_0
\end{pmatrix} \rightarrow ^{\cal D}
{\cal N}(0,  \Sigma_3(\theta_0)) \;\;\mbox{ in distribution under  }
\mathbb{P}_{\theta_0}.
\end{equation}

Using now that the likelihood splits into five terms that can be maximized
separately leads to 
the asymptotic independence of the estimators 
 stated in Proposition \ref{prop:mle}.

\subsection{Proof of Proposition \ref{prop:cond}}\label{appen:cond}
Let us prove by induction Proposition \ref{prop:cond}.
By Assumption A\ref{ST:P}, the property holds for $i=0$ : $S_0$ and
$T_0$ are independent and 
$S_0 \sim
{\cal P}(\sigma)={\cal P}(\Gamma_0)$, 
$T_0 \sim  {\cal P}(\tau )={\cal P}(\Gamma'_0)$.
Assume that the property holds for $i \geq 1$, i.e. 
using definitions  (\ref{def:YGi}), (\ref{def:Gammai}),
$S_i$ and $T_i$ are independent conditionnally on ${\cal G}_{i-1}$ 
and ${\cal L}(S_i/{\cal G}_{i-1}) ={\cal P}(\Gamma_i)$,  
${\cal L}(T_i/{\cal G}_{i-1}) ={\cal P}(\Gamma'_i)$.
Using (\ref{def:S'R'}), 
$
(S_{i+1},R_i)=(S'_{i+1},R'_{i})+ (S''_{i+1},R''_{i}),
$
where the conditional distribution 
of $(S'_{i+1},R'_{i})$ (resp.$(S''_{i+1},R''_{i})$) 
given $(S_i,T_i)$
is the  Multinomial distribution ${\cal M}(S_{i}; a,b)$ (resp. ${\cal M}(T_{i}; a',b')$). 
Using Assumption (A\ref{SiTi}), these two distributions 
are independent conditionally on $ S_i, T_i$
and applying now Lemma \ref{lem:MP} conditionally on $ {\cal G}_{i-1}$
to \{$S_i\sim {\cal P}(\Gamma_{i})$,
${\cal M}(S_{i}; a,b)$\} and to \{$T_i \sim {\cal P}(\Gamma'_{i})$, 
${\cal M}(T_{i}; a',b')$\} 
yields that the four variables $S'_{i+1},R'_{i},S''_{i+1},R'_{i}$  are
independent conditionally on $ {\cal G}_{i-1}$ and that 
$S'_{i+1} \sim {\cal P}(a\Gamma _{i}) $,
$S''_{i+1} \sim {\cal P}(a'\Gamma'_{i-1}) $, 
$R'_i \sim {\cal P}(b\Gamma_{i}) $ and $R''_i \sim {\cal P}(b'\Gamma'_{i}) $.
Hence, $S_{i+1}$ and $R_i$ are independent conditionally on ${\cal G}_{i-1}$
and $S_{i+1} 
\sim {\cal P}(a \Gamma_{i} + a'\Gamma'_{i})$, 
 and $R_i \sim {\cal P}(b\Gamma_{i} + b'\Gamma'_{i})$.

Let us now prove that 
$\mathbb{E}(S_{i+1}/{\cal G}_{i})= \mathbb{E}(S_{i+1}/{\cal G}_{i-1})$.
Let $\phi_1$ and $\phi_2$ two measurable functions of $(Y_{0:i-1},S_{i+1})$
and $(Y_{0:i-1},R_i,V_i,F_i)$ and compute,

$$\mathbb{E}^{{\cal G}_{i-1}}(\phi_1(S_{i+1}) \phi_2(R_i,V_i ,F_i))
= \mathbb{E}^{{\cal G}_{i-1}}( \int \phi_1(s_{i+1}) \phi_2(r_i, v_i,f_i)ds_{i+1}dr_idv_idf_i).
$$

Using (\ref{def:p4p5}) and  (\ref{eq:pitrans}) , set  $\psi_2(r)=  \int \phi_2(r,v,f) p_4(v/r)p_5(f/v)dv 
$, then 
$$\mathbb{E}^{{\cal G}_{i-1}}(\phi_1 (S_{i+1}) \phi_2(R_i,V_i,F_i))
= \mathbb{E}^{{\cal G}_{i-1}}(\phi_1(S_{i+1})\psi_2(R_{i}))$$
$$
= \mathbb{E}^{{\cal G}_{i-1}}(\phi_1(S_{i+1}))\mathbb{E}^{{\cal G}_{i-1}}(\psi_2(R_{i}))= 
\mathbb{E}^{{\cal G}_{i-1}}(\phi_1(S_{i+1})) 
\mathbb{E}^{{\cal G}_{i-1}}(\phi_2(R_{i},V_i,F_i)),
$$
since $S_{i+1}$ and $R_i$ are independent conditionally on ${\cal G}_{i-1}$.
Hence, $S_{i+1}$ and $(R_i,V_i,F_i)$ are independent conditionally on 
${\cal G}_{i-1}$ and, 
$$\mathbb{E}(S_{i+1}/{\cal G}_{i})= \mathbb{E}(S_{i+1}/{\cal G}_{i-1})=
{\cal P}( a\Gamma_{i}+a'\Gamma'_{i+1}).
$$
Consider two measurable functions $\phi_3, \phi_4$ of 
$(Y_{0:i},S_{i+1})$ and $(Y_{0:i},T_{i+1})$, then using 
(\ref{def:p2p3}), (\ref{eq:pitrans}), Assumption
 (A\ref{SiTi}) and the conditional 
independence given ${\cal G}_{i-1}$ of $S_{i+1}$ and  $(R_i,V_i,F_i)$ yield, 
\begin{multline}
\mathbb{E}^{{\cal G}_{i}}(\phi_3(S_{i+1}) \phi_4(T_{i+1}))= 
\mathbb{E}^{{\cal G}_{i}}(\int (\phi_3(s_{i+1})\phi_4(t') p_2(t'/F_i)ds_{i+1}dt')\\
=\mathbb{E}^{{\cal G}_{i}}(\phi_3(S_{i+1}))(\int\phi_4(t')p_2(t/F_i)dt')
= \mathbb{E}^{{\cal G}_{i}}(\phi_3(S_{i+1})) 
\mathbb{E}^{{\cal G}_{i}}(\phi_4(T_{i+1})).
\end{multline}
Therefore, conditionally on  ${\cal G}_{i}$ $S_{i+1}$ and $T_{i+1}$ are
independent and using now Assumption (A\ref{GI:P}), 
${\cal L}(T_{i+1}/{\cal G}_{i})={\cal P}(mF_{i}+u)$. 
The property holds 
for $i+1$ with $\Gamma_{i+1}= a \Gamma_{i} +a'\Gamma'_{i}$ and
 $\Gamma'_{i+1}=mF_{i}+u$, which are the definitions given in (\ref{def:Gammai}).

\subsection{Proof of Theorem \ref{theo:identifiability}}\label{appen:identifiability}
The likelihood $\tilde{l_K^1}(\theta)$ defined in (\ref{def:ltilde1}) sums up 
the available  information 
associated with the incomplete model. An application of the strong law of 
large numbers yields,  
$$
\frac{1}{K}\tilde{l_K^1}(\theta)
\rightarrow L(\theta_0,\theta)=
\mathbb{E}_{\theta _0}\sum_ {i=0}^n (-\Lambda _i(\theta,Y_{0:i-1})
+ R_i\log \Lambda _i(\theta,Y_{0:i-1})). 
$$
According to Theorem \ref{theo:cond}, conditionally
on ${\cal G}_{i-1}$, the random variables $R_i$ are Poisson 
distributions with parameter $\Lambda_i(\theta_0,Y_{0:i-1})$ 
under $\mathbb{P}_{\theta_0}$. Hence, 
$$
L(\theta_0,\theta) = 
\mathbb{E}_{\theta _0} \sum_ {i=0}^n (-\Lambda _i(\theta,Y_{0:i-1})
+\Lambda _i(\theta_0,Y_{0:i-1}) \log \Lambda _i(\theta,Y_{0:i-1})).
$$
Using the explicit form of the Kullback-Leibler divergence 
between two Poisson distributions,
$L(\theta_0,\theta)$ writes,
$$
L(\theta_0,\theta)= - \sum_{i=0}^n 
 \mathbb{E}_{\theta_0}\{{\cal K}({\cal P}(\Lambda_i(\theta_0,Y_{0:i-1})),
{\cal P}(\Lambda_i(\theta,Y_{0:i-1}))) \} 
+ C(\theta_0),$$
with $C(\theta_0)= \mathbb{E}_{\theta_0}\sum_{i=0}^n (-\Lambda_i(\theta_0,Y_{0:i-1})+
\Lambda_i(\theta_0,Y_{0:i-1})\log\Lambda_i(\theta_0,Y_{0:i-1}))$
only depends on the observations.
The identifiability condition for $\theta_0$ 
is therefore equivalent to 
$$
\{L(\theta_0,\theta)=0\}\Rightarrow \{\theta
=\theta_0\}.$$
The Kullback-Leibler divergence of two Poisson distributions ${\cal K}({\cal P}(\mu_0),{\cal P}(\mu)) $
is non negative and equal to 0 if and only if $\mu =\mu_0$.
Hence, the limit $L(\theta_0,\theta)$ presents a strict maximum at
$\theta_0$ 
if and only if 
$$
\Lambda_{i}(\theta,Y_{0:i-1})  = \Lambda_{i}(\theta_0,Y_{0:i-1})
\;\mathbb{P}_{\theta_0}-\mbox{a. s. for  } i=0,\dots, n.
$$
Since the $\Lambda_{i}(\theta,Y_{0:i-1})$ depend on $(F_{i-1},\dots,F_0)$,  
the above condition
can hold only if the coefficients associated with the random variables $F_i$ in the above expression
are equal. The proof below is just elementary algebra based on this
property. We obtain, using for the $c_i(\theta)$ the definitions given
in (\ref{eq:Lambdai}) and (\ref{def:Lambdaik}) :

If $n=0$, $\Lambda_0 (\theta)=c_0(\theta)$ is deterministic. Only 
$ c_0(\theta)$ is identifiable: it is the first
condition stated in Theorem \ref{theo:identifiability}.

If $n=1$, we have, 
$\Lambda_{1}(\theta,Y_{0:i-1})= b'm F_0+c_1(\theta).$
 Since $F_0$ is random, the two random variables
 $\Lambda_{1}(\theta,Y_{0:i-1})$ and   $\Lambda_{1}(\theta_0,Y_{0:i-1})$
are $\mathbb{P}_{\theta_0}$-a.s.  equal iff 
$b'm=b'_0m_0$ and $ c_1(\theta)=c_1(\theta_0)$,
which leads to the identifiability of $ (b'm,
c_0(\theta),c_1(\theta))$.

If $n=2$,  
$\Lambda_2(\theta)= b'm\; F_1+ \frac{a'b}{b'}b'm F_0 +c_2(\theta)$, 
We thus get two additional conditions which lead to  
the identifiability of $(\frac{a'b}{b'},b'm,
c_0(\theta),c_1(\theta),c_2(\theta))$.

If $n=3$, $\Lambda_3(\theta)=  b'm F_2 + \frac{a'b}{b'}b'm (a F_1+F_0)
+c_3(\theta)$. 
Hence, $F_0, F_1,F_2$ being random variables,
 we get  that $(a, \frac{a'b}{b'}, b'm)$ are identifiable.
Now, identifying $(b\sigma,b'\tau,b'u)$ consists in solving 
a linear system using the conditions on 
$c_0(\theta),c_1(\theta),c_2(\theta))$. We obtain :

- if $\{a \neq \frac{a'b}{b'}\} $, $(a,\frac{a'b}{b'},b'm, b'u,
b\sigma,b'\tau)$ is identifiable.

- if $a=\frac{a'b}{b'}$, the identifiable parameters are 
$(a,b'm, b'u, b\sigma+b'\tau)$.

Noting that, for $i\geq 1$, 
$c_{i+1}(\theta)-a c_i(\theta)= (1-a+\frac{a'b}{b'}) b'u$ and that 
only $(a, b'm, \frac{a'b}{b'})$ enters in the $F_i$'s coefficients. 
it can be checked that observing more generations does not lead to 
the idenfiability of
additional parameters.

\subsection{Proof of Proposition \ref{prop:statphi}}\label{appen:statphi}
Let $\phi= (a,\frac{a'b}{b'},b'm,b'u,b\sigma,b'\tau)$
with $a \neq \frac{a'b}{b'}$. Under Asssumption (A\ref{ass:Phi}), we can define $K_1,K_2,a_1,a_2$
such that
$$
\Phi \subset [K_1,K_2]^5 \times [a_1,a_2],\;\mbox{with }
0<K_1<K_2<+\infty \mbox{ and }  0<a_1<a_2<1.
$$ 
The normalized loglikelihood process can be written as 
\begin{equation} 
\frac{1}{K}\tilde{l}_{K}^{\;1}(\phi,Y_{0:n}(K)) = \frac{1}{K}\sum_ {k=1}^K
J(\phi,Y_{0:n}^k)\;\mbox{with }
\end{equation}
$$
J(\phi,Y_{0:n}^k)=\sum_ {i=0}^n R_i^k \log\Lambda _i(\phi,Y_{0:i-1}^k)-
\Lambda _i(\phi,Y_{0:i-1}^k).
$$
These random variables are i.i.d. and  we have to study the behaviour of their
empirical distribution with respect to $\phi$. For getting the consistency 
of the associated maximum likelihood estimator, we have to prove that the
parametric class $\{\mathbb{P}_{\phi}, \phi \in \Phi\}$ 
is Glivenko-Cantelli (see e.g. \citealp{VanderVaart:1998}). 
There is no close form for the density of these variables, so that 
no generic argument can here be applied: we thus propose a direct proof.

The functional $\{\phi \rightarrow J(\phi,Y_{0:n})\}$ is a.s twice 
continuously differentiable on $\Phi$. Let  $D\Lambda_i(\phi)$ denote the gradient in $\mathbb{R}^6$ of
$\Lambda_i(\phi,Y_{0:i-1})$.
Then, 
$$
\mid J(\phi';Y_{0:n})-J(\phi'';Y_{0:n})\mid \leq \biggl(
 \sum_ {i=0}^n R_i \sup_{\phi \in \Phi}
\parallel (\frac{1}{\Lambda_i(\phi)}+1) 
D\Lambda_i(\phi) \parallel  \biggr) \parallel \phi'-\phi''\parallel. 
$$
Under (A\ref{ass:Phi}), we have that, for all $i$, $ \Lambda_i (\phi) \geq K_1>0$.
Let us compute $D\Lambda_i(\phi)$ Using (\ref{def:Lambdai}) for the definitions of
$c_i(\theta)$ and $\Lambda_i(\theta)$, let us set
$$\Lambda_i(\phi)=b'mF_{i-1}+\frac{a'b}{b'}(b'm)(F_{i-2}+aF_{i-3}+\dots+ a ^{i-2}F_0)
+c_i(\phi).$$ 
For $i=0$, we have  
$$\frac{\partial \Lambda_0(\phi)}{\partial a}=
\frac{\partial \Lambda_0(\phi)}{\partial (a'b/b')}=\frac{\partial \Lambda_0(\phi)}{\partial (b'm)}= 
\frac{ \partial\Lambda_0(\phi)}{\partial (b'u)}=0;\;
 \frac{\partial \Lambda_0(\phi)}{\partial (b\sigma)}=
\frac{\partial \Lambda_0(\phi)}{\partial (b'\tau)}=1.
$$
Using the convention that non properly defined terms are set to 0 (e.g. 
the sum $(F_{i-3}+2aF_{i-4}+\dots )$ is set to 0 for $i\leq 2$, we get,
for $i\geq 1$, 
\begin{align*}
\frac{\partial \Lambda_i(\phi)}{\partial a} &=\frac{a'b}{b'}(b'm)
(F_{i-3}+ 2a F_{i-4}+\dots + (i-2)a^{i-3} F_0)+
\frac{\partial c_i(\phi)}{\partial a}, \mbox{ with}\\ 
 \frac{\partial c_i(\phi)}{\partial a} &= (b\sigma)ia^{i-1}+
\frac{a'b}{b'}(b'u)\frac{d}{da}(\frac{1-a^{i-1}}{1-a})
+\frac{a'b}{b'}(b'\tau)(i-1)a^{i-2}\;;\\
\frac{\partial \Lambda_i(\phi)}{\partial (a'b/b')}&=b'm(F_{i-2}+aF_{i-3}+\dots
+a^{i-2}F_0)+  
(b'\tau) a^{i-1} +b'u \frac{1-a^{i-1}}{1-a}\;; \\
\frac{\partial \Lambda_i(\phi)}{\partial (b'm)} &= F_{i-1}
+\frac{a'b}{b'}(F_{i-2}+a F_{i-1}+\dots +a^{i-2}F_0)\;;\\
\frac{ \partial\Lambda_i(\phi)}{\partial (b'u)}&= 1+\frac{a'b}{b'}\frac{1-a^{i-1}}{1-a}\;;\\
\frac{\partial \Lambda_i(\phi)}{\partial (b\sigma)}& = a^i\;;\\
\frac{\partial \Lambda_i(\phi)}{\partial (b'\tau)}& =\frac{a'b}{b'}a^{i-1}.
\end{align*}
All these partial derivatives are positive for $i\geq 1$ and, except the first one 
$\frac{\partial \Lambda_i(\phi)}{\partial a}$, 
they are bounded from above by
$M_1 (1+\sum_{i=0}^{i-1}F_j)$ where $M_1$ is a constant determined by
$\Phi$ (since on $\Phi$, $0 <a_1 \leq a\leq a_2 <1$).
Noting now that the application $\{i\rightarrow ia^{i-1}\}$ satisfies
 $ \; \forall i\geq 1,\; 
 \alpha_1<ia^{i-1}<\alpha_2$ on $[a_1,a_2]\subset (0,1)$ with 
$0<\alpha_1<\alpha_2<+\infty$, we can also bound
$\frac{\partial \Lambda_i(\phi)}{\partial a}$ 
by  $ M_2 (1+ \sum_{j=0}^{i-3}F_j)$. Joining these bounds, we  get, 
$$
\sup \parallel (\frac{1}{\Lambda_i(\phi)}+1) D\Lambda_i(\phi) \parallel 
\leq M_3(1+\sum_{j=0}^{i-1} F_j )\;,$$
$$
\mid J(\phi;Y_{0:n})-J(\phi';Y_{0:n})\mid \leq \eta M_3 Z_n \mbox{ with }
Z_n =\sum_ {i=0}^n (1+ \sum_{j=0}^{i-1}F_j)R_i.
$$
Using now that
$\mathbb{E}_{\phi_0}(R_i \sum_{j=0}^{i-1} F_j)=
\mathbb{E}_{\phi_0}(\Lambda_i(\phi_0) \sum_{j=0}^{i-1} F_j)$, $Z_n$ satisfies, 
$$\mathbb{E}_{\phi_0}Z_n 
\leq n \mathbb{E}_{\phi_0}\biggl(\sum_ {i=0}^n (1+ F_j)^2 \biggr)
\leq n \sum_ {i=0}^n \biggl(1+ 
\sqrt{\mathbb{E}_{\phi_0}F_i^2}
\biggr)^{2}.$$
This is finite since $G(.)$ and $\mu(.)$ have finite variance and $n$ is prescribed.\\
Let us define the r.v. $Z_n^k= \sum_ {i=0}^n (1+ \sum_{j=0}^{i-1}F_j^k)R_i^k$. 
Then the continuity modulus of  $\tilde{l}^1_K(\theta)$ verifies : 
$$
w(K,\eta,\frac{1}{K}\tilde{l}_K^1) = \sup_{\parallel\phi'-\phi'' \parallel 
\leq \eta }  \frac{1}{K}\sum_{k=1}^K \mid
J(\phi';Y_{0:n}^k) -J(\phi'';Y_{0:n}^k) \mid \;
\leq \eta \frac{1}{K}\sum_{k=1}^K  Z_n^k.
$$
Using now that $\mathbb{E}_{\phi_0}Z_n< \infty$, we can apply the strong law of large numbers to 
$w(K,\eta,\frac{1}{K}\tilde{l}_K^1)$, which is a sufficient condition for ensuring the 
consistency 
of $\hat{\phi}_K$.

It remains to study the asymptotic normality of the estimators.
It is easy to check that the random variables $^{\tau}W^k=(W_1^k,\dots,W_6^k)$ with 
$$W_p^k= \sum _{i=0}^n(\frac{R_i^k}{\Lambda_i^k(\phi_0)}-1)
\frac{\partial \Lambda_i^k(\phi_0)}{\partial \phi_p}$$
are i.i.d. centered under $\mathbb{P}_{\phi_0}$,
with covariance matrix $\Sigma$ is for $1\leq p,q\leq 6$
$$
\Sigma_{p,q}=\Sigma_{p,q} (\phi_0)=\mathbb{E}_{\phi_0}
\sum_{i=0}^{n}\frac{1}{\Lambda_i(\phi_0)}\,
\frac{\partial \Lambda_i}{\partial \phi_p}(\phi_0)\,
\frac{\partial \Lambda_i}{\partial \phi_q}(\phi_0).$$
This matrix is well-defined and finite is since, 
for all $i$, $\Lambda_i (\phi) \geq K_1 >0 $ and  
$\mid \frac{\partial \Lambda_i(\phi)}{\partial \phi_p }\mid
 \leq M_3(1+ \sum_0 ^{i-1}F_j)$ for $p=1,\dots 6$.
Now, $\hat{\phi}_K $ is a zero for the score function, so that a Taylor expansion at 
$\phi_0$ yields,  using the consistency of the vector $\hat{\phi}_K$, 
$$
0=\frac{1}{\sqrt K}D\tilde{l}^1_K(\hat{\phi}_K)=\frac{1}{\sqrt K}
D\tilde{l}^1_K(\phi_0) +
\frac{1}{K} \biggl(\nabla \tilde{l}^1_K(\phi_0)+ R_K(\phi_0,\hat{\phi}_K\biggr)
\sqrt K (\hat{\phi}_K-\phi_0) $$
The term $\nabla \tilde{l}^1_K(\phi_0)$ contains the second derivatives 
of $\tilde{l}^1_K(\phi)$ w.r.t. $\phi_p,\phi_q$ and
$\tilde{R}_K(\phi_0,\hat{\phi}_K)$ 
is the remainder
term of the Taylor expansion.  
So, the strong law of large numbers yields :
$\frac{1}{K}\nabla \tilde{l}^1_K(\phi_0) \rightarrow -
\sum_ {i=0}^n\mathbb{E}_{\phi_0}(
\frac{1}{\Lambda_i(\phi_0)}\frac{\partial \Lambda_i(\phi_0)}
 {\partial \phi_p}\frac{\partial \Lambda_i(\phi_0)}{\partial \phi_q})
 $.
The remainder term $R_K(\phi_0,\hat{\phi}_K) $ is bounded uniformly on $\Phi$
by $\parallel \hat{\phi}_K-\phi_0 \parallel Z_K$,
with $Z_K= sup\{\frac{1}{K} \nabla \tilde{l}^1_K(\phi), \phi \in \Phi \}$. Using 
that $Z_K$
is bounded uniformly on
$\Phi$ by  $ n^2 M_6(1 +\sum_{i=0}^{n}F_i F_j)^2$, we get that
$R_K(\phi_0,\hat{\phi})$ goes to $0$ under $\mathbb{P}_{\phi_0}$, 
which leads to the result stated in Proposition \ref{prop:statphi} provided
that $\Sigma(\phi_0)$ is invertible.

\bigskip

{\bf Aknowledgments}\\
This study was funded by the European project "sustainable Introduction of Genetically Modified Crops into 
European Agriculture" and by the project "Flux des (trans-)g\`enes et impact sur la biodiversit\'e" of the
Agence nationale de la Recherche. 
\bibliographystyle{biom}
\bibliography{AOC}

\end{document}